\pdfoutput=1
\documentclass[11pt,twoside,a4paper,cmspaper,final,collab]{cms-tdr}
\def\svnVersion{fd79b8d}\def\svnDate{2026/04/24}\def\cmsCernNoTag{CERN-EP-2025-129}\def\cmsCernDate{\today}\def\cmsMessage{Published in the Journal of High Energy Physics as \href{http://dx.doi.org/10.1007/JHEP04(2026)132}{\doi{10.1007/JHEP04(2026)132}.}}
\begin{document}\cmsNoteHeader{SUS-24-002}

\newcommand{\maO}{\ensuremath{m_{\Pa_1}}\xspace}
\newcommand{\PaO}{\ensuremath{\HepParticle{\Pa}{1}{}}\xspace}
\newcommand{\PgPg}{\ensuremath{{\Pg\Pg}\text{F}}\xspace}
\newcommand{\SR}{\ensuremath{\textit{SR}}\xspace}
\newcommand{\CR}{\ensuremath{\textit{CR}}\xspace}
\newcommand{\CRNtwothree}{\ensuremath{\CR\: N_{23}}\xspace}
\newcommand{\CRNisoone}{\ensuremath{\CR\: N_{{\text{iso}},1}}\xspace}
\newcommand{\CRNisotwothree}{\ensuremath{\CR\: N_{{\text{iso}},23}}\xspace}
\newcommand{\CRLooseIso}{\ensuremath{\textit{CR Loose-Iso}}\xspace}

\cmsNoteHeader{SUS-24-002}

\title{Search for light pseudoscalar boson pairs produced from Higgs boson decays using the \texorpdfstring{$4\PGt$}{4tau} and \texorpdfstring{$2\PGm2\PGt$}{2mu2tau} final states in proton-proton collisions at \texorpdfstring{$\sqrt{s}=13\TeV$}{sqrt(s) = 13 TeV}}

\date{\today}

\abstract{
A search for a pair of light pseudoscalar bosons (\PaO) produced in the decay of the 125\GeV Higgs boson is presented. The analysis examines decay modes where one \PaO decays into a pair of tau leptons and the other decays into either another pair of tau leptons or a pair of muons. The \PaO boson mass probed in this study ranges from 4 to 15\GeV. The data sample was recorded by the CMS experiment in proton-proton collisions at a center-of-mass energy of 13\TeV and corresponds to an integrated luminosity of 138\fbinv. No excess above standard model (SM) expectations is observed. The study combines the $4\PGt$ and $2\PGm2\PGt$ channels to set upper limits at 95\% confidence level (\CL) on the product of the Higgs boson production cross section and the branching fraction to the $4\PGt$ final state, relative to the Higgs boson production cross section predicted by the SM. In this interpretation, the \PaO boson is assumed to have Yukawa-like couplings to fermions, with coupling strengths proportional to the respective fermion masses. The observed (expected) upper limits range between 0.007 (0.011) and 0.079 (0.066) across the mass range considered. The results are also interpreted in the context of models with two Higgs doublets and an additional complex singlet field (2HD+S). The tightest constraints are obtained for the Type~III 2HD+S model. In this case, assuming the Higgs boson production cross section equals the SM prediction, values of the branching ratio for the Higgs boson decay into a pair of \PaO bosons exceeding 16\% are excluded at 95\% \CL for \PaO boson masses between 5 and 15\GeV and $\tan\beta > 2$, with the exception of scenarios in which the \PaO boson mixes with charm or bottom quark-antiquark bound states.
}

\hypersetup{%
pdfauthor={CMS Collaboration},%
pdftitle={Search for light pseudoscalar boson pairs produced from Higgs boson decays using the 4tau and 2mu2tau final states in proton-proton collisions at sqrts = 13 TeV},%
pdfsubject={CMS},%
pdfkeywords={CMS, Higgs, pseudoscalars, 2HD+S, muons, taus}}

\maketitle

\section{Introduction}
\label{Sec:introduction}

The experimental verification of the standard model (SM), a quantum field theory describing the strong and electroweak interactions between elementary particles, culminated in the discovery of a scalar boson with a mass of 125\GeV by the ATLAS and CMS Collaborations at the CERN LHC~\cite{Aad:2012tfa, Chatrchyan:2012xdj, CMS:2013btf}. Subsequent studies demonstrated that the properties of the discovered boson, denoted henceforth as \PH, are compatible, at the current level of precision, with the predicted properties of the SM Higgs boson~\cite{ATLAS:2022vkf, CMS:2022dwd}. 

Despite its striking success in describing the subatomic world, the SM is not considered a complete theory. It does not explain several key phenomena, such as the nature of dark matter, the origin of neutrino masses, and the matter-antimatter asymmetry in the universe. Various extensions to the SM have been proposed to address some of its shortcomings, which include extended Higgs sectors wherein the 125\GeV Higgs boson appears as one of multiple Higgs boson states. Prominent examples of such extensions of the SM are the general two-Higgs-doublet models~\cite{Lee:1973iz, Branco:2011iw}, and the two-Higgs-doublet plus singlet (2HD+S) models, which include an additional complex scalar field~\cite{Curtin:2013fra}. Extended Higgs sectors are also embedded in more elaborate theoretical frameworks, such as the minimal supersymmetric SM (MSSM)~\cite{Fayet:1974pd, Fayet:1977yc}, which incorporates two Higgs doublets, or the next-to-MSSM (NMSSM)~\cite{Ellwanger:2009dp, Maniatis:2009re}, which solves the $\mu$-problem of the MSSM by extending the two Higgs doublets by a complex singlet field~\cite{Kim:1983dt,Ramos-Sanchez:2010wxr}. These supersymmetric extensions of the SM provide a candidate for dark matter and introduce additional sources of charge-parity (CP) violation that could explain the observed amount of matter-antimatter asymmetry. 

The analysis presented in this paper probes the 2HD+S models, which predict seven physical Higgs boson states: three neutral CP-even, two neutral CP-odd, and two charged bosons. One of the scalars of the neutral sector can be identified as the 125\GeV Higgs boson. At the same time, one of the two pseudoscalars, denoted \PaO, can be light enough for $\PH\to\PaO\PaO$ decays to be kinematically possible. Measurements of the Higgs boson couplings so far still allow for a substantial branching fraction of Higgs boson decays to beyond-the-SM particles; the ATLAS and CMS experiments have established upper limits at 95\% confidence level (\CL) of 12 and 16\%, respectively~\cite{ATLAS:2022vkf, CMS:2022dwd}. The search for $\PH\to\PaO\PaO$ decays, therefore, provides an effective probe for possible extensions of the SM and the discovery of new physics phenomena.

The 2HD+S models predict various decay channels for the \PaO boson, with significant branching fractions into fermion pairs. The decay patterns of \PaO bosons are governed by parameters such as the ratio of the vacuum expectation values of the two Higgs doublets ($\tan\beta$) and the \PaO boson mass (\maO). The different variants of the 2HD+S models---Type~I, II, III, and IV---are distinguished by how the Higgs doublets couple to fermions, resulting in distinct decay patterns. Several of these models predict a significant branching fraction of the \PaO boson decay into a pair of tau leptons~\cite{Curtin:2013fra}, making this channel a promising avenue for experimental searches. 

Numerous searches for $\PH\to\PaO\PaO$ decays have been carried out to date by the ATLAS and CMS Collaborations, exploring different \PaO decay modes, covering the mass range 0.2 $< \maO <$ 62.5\GeV~\cite{ATLAS:2015hpr, CMS:2012qms, CMS:2015nay, CMS:2017dmg, CMS:2015twz, ATLAS:2015rsn, ATLAS:2018coo, CMS:2018jid, ATLAS:2015unc, CMS:2018qvj, CMS:2020ffa, ATLAS:2018emt, CMS:2024uru, ATLAS:2018jnf, CMS:2019spf, CMS:2024zfv, CMS:2022xxa, CMS:2022fyt, ATLAS:2025qyn}. The studies did not reveal significant deviations from the SM background expectations, and upper limits were set on the signal rates, thereby constraining the parameters of the 2HD+S models.      

In this paper, a search for a light \PaO boson via the decays $\PH\to\PaO\PaO\to 4\PGt$ and $\PH \to \PaO\PaO \to 2\PGm 2\PGt$ is presented. The search is based on proton-proton ($\Pp\Pp$) collision data at a center-of-mass energy of 13\TeV, corresponding to an integrated luminosity of 138\fbinv, recorded by the CMS detector between 2016 and 2018. The analysis covers the \maO range between 4 and 15\GeV and employs a specialized strategy to select and identify highly Lorentz-boosted muon or tau lepton pairs with overlapping decay products. Similar searches were performed by the CMS and ATLAS Collaborations~\cite{CMS:2019spf,ATLAS:2025qyn} using 36 and 140\fbinv of $\Pp\Pp$ collision data, respectively. These searches established upper limits at 95\% \CL on the Higgs boson production cross section times branching fraction for $\PH\to\PaO\PaO\to4\PGt$, relative to the SM Higgs boson production cross section. The CMS analysis set limits in the range 0.022--0.23, depending on the \PaO boson mass. The ATLAS analysis, based on a larger analyzed data set, placed limits between 0.03 and 0.10. The analysis presented here extends the previous CMS analysis by using approximately four times more data. Moreover, the signal selection criteria have been optimized further, taking advantage of a more precise calibration of the CMS detector achieved towards the end of the LHC Run 2 data-taking period. 

The $\PH\to\PaO\PaO\to 4\PGt$ decay gives rise to tau leptons that have a lower average transverse momentum (\pt) spectrum compared to the $\PH\to\PGt\PGt$ decay. For about 60\% of tau leptons from the signal processes, the visible \pt, defined as the transverse momentum of its detectable decay products, is below 20\GeV. Reconstruction and identification of tau leptons in this kinematic regime is challenging. In particular, the reconstruction of decay modes with neutral pions is hampered by a large flux of low-\pt photons originating from low-energy $\Pp\Pp$ interactions in the same or neighboring bunch crossings (pileup). Conventional CMS algorithms~\cite{CMS:2022prd} are not optimized for identifying such low-\pt tau leptons, resulting in a rapid degradation of the reconstruction efficiency and purity with decreasing visible \pt. Furthermore, the significant mass difference between the \PaO and Higgs bosons results in \PaO bosons that are highly Lorentz-boosted, and collimated decay products. The resulting overlap of tau leptons from the same \PaO boson further complicates the identification of the $\PaO\to\PGt\PGt$ candidates. These challenges are addressed by employing a dedicated search strategy. 

\begin{figure}[t]{
\centering
\includegraphics[width=0.7\textwidth]{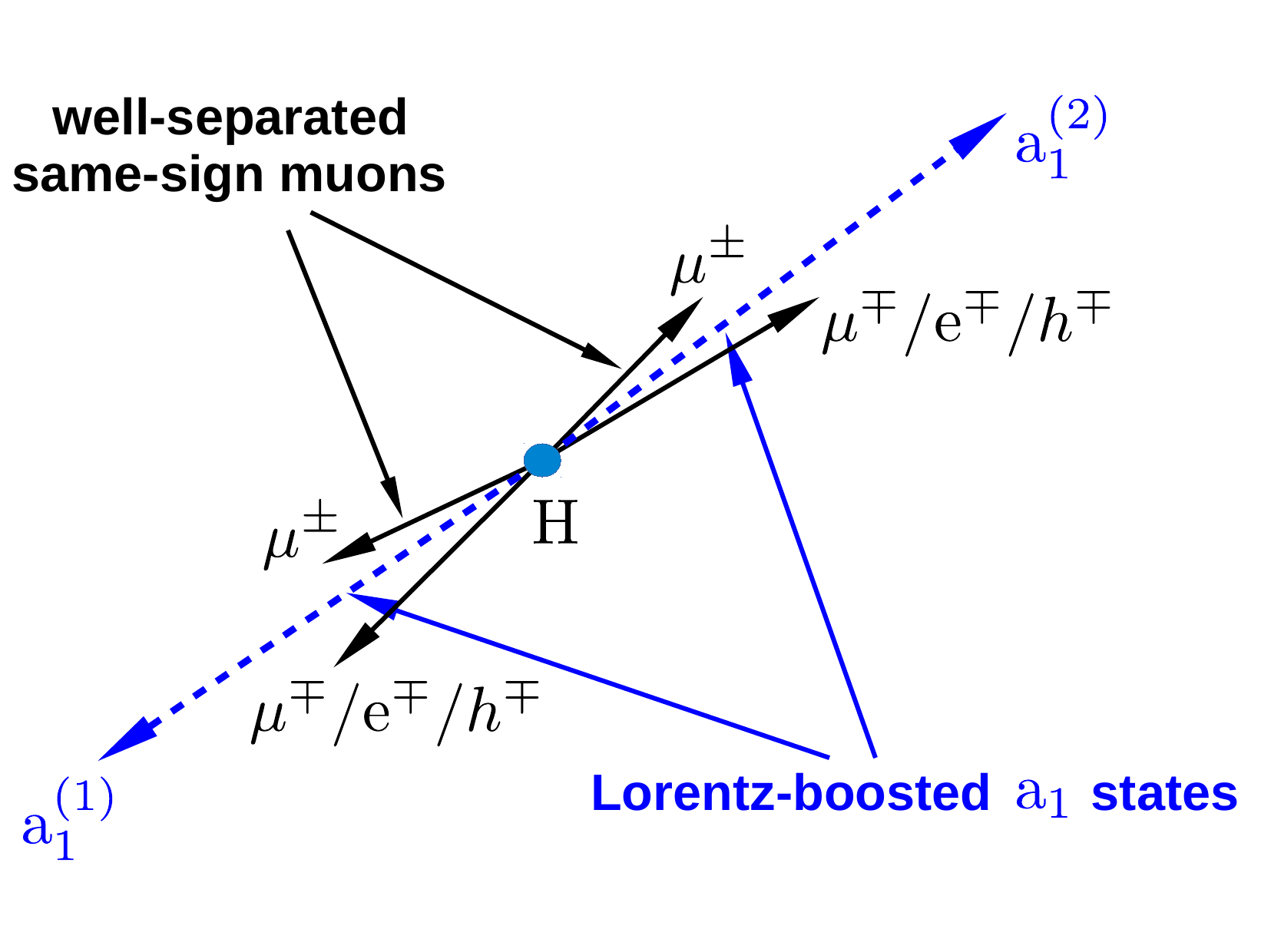}
\caption{
Illustration of the signal topology. The Higgs boson decays into two \PaO bosons, one of which further decays into a pair of tau leptons, and the other into a pair of muons or tau leptons. In the case of $\PaO \to \PGt \PGt$ decays, one tau lepton is required to decay to a muon, and the other to a single charged particle---either an electron, a muon, or a charged hadron ($h$). The targeted final state consists of one muon and one oppositely charged track arising from each \PaO boson decay.
\label{fig:topology}
}
}
\end{figure}

The signal topology targeted by the present analysis is illustrated in Fig.~\ref{fig:topology}. The analysis focuses on the $\PH \to \PaO\PaO\to(\PGt_{\PGm}\PGt_{\text{1-prong}})$ $(\PGt_{\PGm}\PGt_{\text{1-prong}})$ decays, where $\PGt_{\PGm}$ denotes the muonic decay of a tau lepton, and $\PGt_{\text{1-prong}}$ stands for its decay into an electron, muon, or one-prong hadronic state. Each \PaO boson is thus identified by the presence of a muon and an additional charged particle nearby. The latter particle is characterized by the presence of one reconstructed track with a charge opposite to that of the muon. Three-prong tau decays are excluded because of their very high quantum chromodynamics (QCD) multijet background and lower reconstruction efficiency. Although the analysis primarily targets the $(\PGt_{\PGm}\PGt_{\text{1-prong}})(\PGt_{\PGm}\PGt_{\text{1-prong}})$ decays, the decays $\PaO\PaO\to(\PGm\PGm)(\PGt_{\PGm}\PGt_{\text{1-prong}})$ are also included as they can give rise to a similar topology. The two decay modes are combined to search for the $\PH \to \PaO\PaO$ signal, assuming that the \PaO boson has Yukawa-like couplings to fermions, with coupling strengths proportional to the respective fermion masses. Because of its significantly lower branching fraction, the $\PaO\PaO\to 4\PGm$ decay has a negligible contribution to the search sensitivity and is not considered in the present search. 

The analysis primarily targets the dominant gluon-gluon fusion (\PgPg) production mechanism, where the Higgs boson is produced with relatively small \pt and the \PaO bosons are emitted nearly back-to-back in the transverse plane, resulting in a large separation in azimuthal angle ($\Delta\phi$) between the decay products of the two \PaO bosons. When the Higgs boson is produced with high \pt, \eg, due to initial-state radiation in \PgPg, or through other production mechanisms, the azimuthal angle between the \PaO bosons is reduced, whereas the separation in pseudorapidity ($\Delta\eta$) can still be large. Therefore, the analysis focuses on the identification of same-sign dimuon events with significant angular separation, $\Delta R = \sqrt{\smash[b]{(\Delta\phi)^{2}+(\Delta\eta)^2}}$, where each muon is accompanied by a nearby oppositely charged particle from the same \PaO decay. The requirement of same-sign muons significantly reduces backgrounds from top quark pairs, Drell--Yan processes, and diboson production, thereby enhancing the search sensitivity. 

Outside the probed mass range of 4--15\GeV, the sensitivity of the present search decreases rapidly. At lower \PaO masses, the background from QCD multijet events overwhelms the signal. At higher masses, the angular separation between the decay products of the \PaO bosons increases, thereby necessitating a different analysis strategy.    

It should be emphasized that the analysis strategy, background modeling, and binning of the final discriminant are devised to target the $4\PGt$ component of the signal. The $2\PGm2\PGt$ channel is added as a supplementary signal component within the same framework. A study aimed specifically at the $2\PGm2\PGt$ final state would require a dedicated strategy, involving the search for a resonant structure in the invariant mass of the $\PaO\to\PGm\PGm$ candidate~\cite{ATLAS:2015unc,CMS:2018qvj,CMS:2020ffa}.

Tabulated results are provided in the HEPData record for this analysis~\cite{hepdata}.

\section{The CMS detector}
\label{Sec:detector}
The central feature of the CMS apparatus is a superconducting solenoid of 6\unit{m} internal diameter, providing a magnetic field of 3.8\unit{T}. Within the solenoid volume are a silicon pixel and strip tracker, a lead tungstate crystal electromagnetic calorimeter (ECAL), and a brass and scintillator hadron calorimeter, each composed of a barrel and two endcap sections. Forward calorimeters extend the pseudorapidity coverage provided by the barrel and endcap detectors. Muons are reconstructed in gas-ionization detectors embedded in the steel flux-return yoke outside the solenoid. More detailed descriptions of the CMS detector, together with a definition of the coordinate system used and the relevant kinematic variables, can be found in Refs.~\cite{CMS:2008xjf, CMS:2023gfb}.

Events of interest are selected using a two-tiered trigger system. The first level (L1), composed of custom hardware processors, uses information from the calorimeters and muon detectors to select events at a rate of around 100\unit{kHz} within a fixed latency of about 4\mus~\cite{CMS:2020cmk}. The second level, known as the high-level trigger, consists of a farm of processors running a version of the full event reconstruction software optimized for fast processing, which reduces the event rate to around 1\unit{kHz} before data storage~\cite{CMS:2016ngn, CMS:2024aqx}.

\section{Simulated samples}
\label{Sec:simsamples}

The signal events for the $\PH \to \PaO \PaO \to 4\PGt$ channel are simulated at leading order (LO) using the \PYTHIA (v.8.212) event generator~\cite{Sjostrand:2014zea}, targeting three major Higgs boson production mechanisms: \PgPg, vector boson fusion (VBF), and Higgs-strahlung (VH). For the $\PH \to \PaO \PaO \to 2\PGm 2\PGt$ decay channel, events are generated at LO for the dominant \PgPg production mode with \MGvATNLO (v.2.6.5)~\cite{Alwall:2014hca}. The acceptance of the $\PH \to \PaO \PaO \to 2\PGm 2\PGt$ signal is corrected to account for subdominant contributions from the VBF and VH processes as detailed in Section~\ref{Sec:selection}. Simulation studies have demonstrated that the signal acceptance is highly sensitive to the \pt spectrum of the Higgs boson. Therefore, the signal samples generated at LO are reweighted to match the Higgs boson \pt spectrum predicted by theoretical computations that incorporate higher-order corrections. The \pt distribution of the Higgs boson produced via \PgPg is reweighted for both decay channels ($4\PGt$ and $2\PGm2\PGt$) using $K$-factors calculated with the program HqT (v.2.0)~\cite{Bozzi:2005wk, deFlorian:2011xf}, which computes the \pt spectrum of the Higgs boson at next-to-LO (NLO) with next-to-next-to-leading-logarithmic resummation at low \pt. For VBF and VH processes, the \pt distribution of the Higgs boson is reweighted using $K$-factors calculated with the NLO \textsc{PowhegBox} (v.2.0) event generator~\cite{Alioli:2008tz, Nason:2004rx, Frixione:2007vw}.

The major source of background for the analysis is QCD multijet production, followed by subdominant contributions from Drell--Yan processes with \PZ or \PW boson production accompanied by jets ($\PZ$+jets and $\PW$+jets), top quark-antiquark pair production with additional jets ($\ttbar$+jets), single top quark production, and vector boson pair production (diboson). The QCD multijet and diboson backgrounds are simulated using \PYTHIA (v.8.212), whereas the $\PZ$+jets and $\PW$+jets backgrounds are generated at LO with \MGvATNLO (v.2.6.5). The single top quark and $\ttbar$+jets processes are generated at NLO using \textsc{PowhegBox} (v.2.0). 

Parton showering and fragmentation for all Monte Carlo (MC) samples are executed using \PYTHIA (v.8.212). The CP5 tune is applied to describe the underlying event~\cite{CMS:2019csb}. All simulated samples utilize the NNLO NNPDF3.1 PDFs. For samples produced at LO with \MGvATNLO, the MLM jet matching scheme is employed~\cite{Alwall:2007fs}. The detector response is simulated with the \GEANTfour package~\cite{GEANT4:2002zbu, Allison:2006ve}. The contribution of pileup is replicated by simulation of minimum bias interactions, which are then superimposed on the primary hard-scattering event. The simulated events are then reweighted to reflect the observed pileup distribution in the experimental data. The average number of pileup collisions was 27 (38) in the 2016 (2017--2018) data. 

\section{Event reconstruction}
\label{Sec:reconstruction}

Events containing two muon candidates and two additional tracks are selected in the analysis. Track reconstruction is performed using the combinatorial track finder algorithm~\cite{CMS:2014pgm}, which employs Kalman filtering to refine the track estimates; the analysis selects the ``high-purity'' tracks defined in Section 4.4 of Ref.~\cite{CMS:2014pgm}. 

The primary vertex (PV) is identified as the vertex corresponding to the hardest scattering in the event, evaluated using tracking information alone, as described in Section 9.4.1 of Ref.~\cite{CMS-TDR-15-02}.

The particle-flow (PF) algorithm~\cite{CMS:2017yfk} reconstructs and identifies individual particles (PF candidates) in an event using combined information from the various elements of the CMS detector. Muons are identified as tracks in the central tracker consistent with either a track or several hits in the muon system and associated with calorimeter deposits compatible with the expectation for minimum-ionizing particles. These PF muons are further required to pass dedicated identification requirements, which depend on several parameters, including track quality, impact parameter significance, and number of muon chamber hits, to suppress contributions from nonprompt decays of hadrons into muons and their punch\-through to the muon detectors. The analysis employs the medium identification criterion~\cite{CMS:2018rym}, which yields an overall efficiency of 98--99\% for muons with $\pt > 20\GeV$. Scale factors, derived by comparing the muon identification efficiencies between data and simulation, are applied to the MC samples to match the performance observed in data. 

Jets are reconstructed by clustering PF candidates using the anti-\kt algorithm~\cite{Cacciari:2008gp, Cacciari:2011ma} with a distance parameter of 0.4. Charged particles not associated with the PV are excluded from the clustering using the charged hadron subtraction method~\cite{CMS:2017yfk}. The reconstructed jet energies are corrected in both data and simulation to account for effects from nonlinear detector response and contamination from pileup~\cite{CMS:2016lmd}. The jet energy resolution is corrected in simulation to improve the agreement with data. 

Jets resulting from the fragmentation and hadronization of bottom quarks (\PQb jets) are identified using the \textsc{DeepJet} algorithm~\cite{Bols:2020bkb, CMS-DP-2023-005}, and labeled as \PQb-tagged. The tight working point of the \PQb tagging discriminator is used, which corresponds to a rate of 0.1\% for misidentifying a light jet, \ie, a gluon or light-quark jet, as a \PQb jet~\cite{CMS:2017wtu}. The corresponding \PQb tagging efficiency, measured from $\ttbar$+jets events, is around 65\%. 

\section{Event selection}
\label{Sec:selection}
The search, targeting nonisolated same-sign muon pairs, requires a specialized triggering strategy. Events of interest are recorded using a set of dimuon triggers. In 2016 and 2017, the trigger required the highest-\pt (leading) and second-highest-\pt (subleading) muons to have transverse momenta of at least 17 and 8\GeV, respectively. In 2018, the thresholds were raised to 18~and 9\GeV, respectively. Additionally, in 2016 and 2018, the trigger required the two muons to have the same charge. In 2016, it was also required that the two muon tracks have their points of closest approach to the beam axis within 0.2\cm of each other along the $z$ direction. The triggers employed in 2016 and 2018 did not impose any isolation requirements on the muons. However, in 2017, a nonisolated same-sign dimuon trigger was not available. Thus, a dimuon trigger imposing a loose isolation criterion on both muons was used instead. This criterion required the ratio of the \pt sum of charged hadrons, within an isolation cone of size $\Delta R = 0.3$ around the muon, to the \pt of the muon to be less than 0.4. Even with this requirement, the trigger was efficient in selecting the signal with a relative loss of efficiency, estimated from simulation studies, ranging from 5\% at $\maO=15\GeV$ to 20\% at $\maO=4\GeV$.    

As the first step in the offline selection, a \PQb jet veto is imposed, rejecting events that contain one or more \PQb-tagged jets with $\pt > 20\GeV$ and $\abs{\eta} < 2.4$. This veto is essential to suppress background from QCD multijet events with heavy-flavor hadrons that decay into muons. The remaining events that pass the trigger selection criteria are required to contain at least two same-sign muons, each having $\abs{\eta} < 2.4$. The \pt of the leading and subleading muon is required to be greater than 19 and 10\GeV, respectively. The transverse (longitudinal) impact parameter of the muons with respect to the PV is required to be $\abs{d_0}<0.05$ ($\abs{d_z}<0.1$)\unit{cm}. The angular separation between the muons must be $\Delta R>1.5$. If more than one pair of same-sign muons meets these criteria, the pair with the highest scalar sum of \pt is selected.

The next selection step employs information from the tracks associated with the reconstructed charged PF objects, excluding the pair of same-sign muons. This information is used to identify and isolate the candidates for the $\PaO\to \PGt_\PGm\PGt_{\text{1-prong}}$ or $\PaO\to\PGm\PGm$ decays, hereafter referred to as \PaO candidates. Two types of tracks are considered, namely ``isolation'' and ``signal'' tracks. The signal tracks are a subset of the isolation tracks, subject to stricter selection criteria. The purpose of each track type and their selection criteria are given in Table~\ref{tab:tracktypes}.

\begin{table}[htbp!]
\centering
\topcaption{The purpose and selection criteria for the two types of tracks used in the selection procedure.}
\begin{tabular}{lccccp{5cm}}
Type of track & $\pt$ & $\abs{\eta}$ & $\abs{d_0}$ & $\abs{d_z}$ & Purpose \\
\hline
 & & & & & \\
Isolation & $>$1.0\GeV & $<$2.4 &  $\phantom{0}<$0.2\cm & $\phantom{0}<$0.3\cm  & Provides the isolation criteria for muon candidates \\
& & & & & \\
Signal    & $>$2.5\GeV & $<$2.4 & $<$0.02\cm & $<$0.04\cm & Used together with muons to construct \PaO candidates \\
\end{tabular}
\label{tab:tracktypes}
\end{table}

Each selected muon of the same-sign muon pair is required to have exactly one isolation track within a $\Delta R$ cone of 0.5 around the muon. The background components, especially QCD multijet events, tend to have higher track multiplicity with respect to the signal and, hence, are rejected by imposing the isolation requirement. A muon+track system is accepted as an \PaO~candidate if the isolation track around the muon meets the signal track criteria. The event is selected in the final sample if it contains two \PaO~candidates. The set of selection requirements outlined above defines the signal region (\SR).

The number of observed events selected in the \SR is 7803. Simulation studies indicate that QCD multijet events are the predominant background in the \SR. Contributions from other backgrounds only comprise about $1\%$ of the selected events in the \SR. In Table~\ref{tab:signal_selection}, the expected signal acceptance times selection efficiency (${\mathcal{A}}\epsilon$), and the expected yield of signal events in the \SR from simulation are reported for a few representative values of \maO. For each of the two considered decay channels, $4\PGt$ and $2\PGm2\PGt$, the value of ${\mathcal{A}}\epsilon$ is evaluated as the expected yield of signal events from simulation relative to the total number of signal events expected in the respective decay channel. The expected signal yields are calculated assuming a benchmark value of the branching fraction, ${\mathcal{B}}(\PH \to \PaO \PaO) {\mathcal{B}}^{2} (\PaO \to \PGt \PGt)=0.05$, and the SM predictions for the Higgs boson production cross sections: 48.58\unit{pb} for \PgPg, 3.78\unit{pb} for VBF, and 2.26\unit{pb} for VH~\cite{deFlorian:2016spz}. The contributions from the \PgPg, VBF, and VH processes are summed to determine the $4\PGt$ yield. The expected $2\PGm2\PGt$ signal yield is estimated under the assumption that the \PaO boson has Yukawa-like couplings to fermions, with coupling strengths proportional to the fermion masses. Under this assumption, the ratio of the $\PaO\to\PGm\PGm$ and $\PaO\to\PGt\PGt$ partial widths satisfies the relation~\cite{aDecays}
\begin{equation}
\frac{\Gamma(\PaO \to \PGm\PGm)}{\Gamma(\PaO \to \PGt\PGt)}
=\frac{m^2_{\PGm}}{m^2_{\PGt}} \frac{\sqrt{1-\big(2m_{\PGm}/\maO\big)^2}}{\sqrt{1-\big(2m_{\PGt}/\maO\big)^2}}.
\label{eq:width_ratios}
\end{equation}

Using Eq.~(\ref{eq:width_ratios}), the ratio of the branching fractions for $\PaO\PaO \to 2\PGm 2\PGt$ and $\PaO\PaO \to 4\PGt$ decays is computed as
\begin{equation}
\frac{{\mathcal{B}}(\PaO\PaO \to 2\PGm 2\PGt)}{{\mathcal{B}}(\PaO\PaO \to 4\PGt)}= 2\frac{{\mathcal{B}}(\PaO \to \PGm\PGm)}{{\mathcal{B}}(\PaO \to \PGt\PGt)}= 2\frac{\Gamma(\PaO \to \PGm\PGm)}{\Gamma(\PaO \to \PGt\PGt)}.
\label{eq:br_ratios}
\end{equation}

The factor of 2 in Eq.~(\ref{eq:br_ratios}) accounts for the possibility of either \PaO boson decaying into $\PGm\PGm$ or $\PGt\PGt$. The value of the ratio in Eq.~(\ref{eq:br_ratios}) decreases from 0.016 at $\maO=4\GeV$ to 0.0073 at $\maO=15\GeV$. To account for subdominant contributions to the $2\PGm2\PGt$ final state from VBF and VH production, the ratios of ${\mathcal{A}}\epsilon$ between the $2\PGm2\PGt$ and $4\PGt$ channels are assumed to be identical across all production mechanisms. The value of ${\mathcal{A}}\epsilon$ for the \PgPg process is then scaled accordingly to estimate the total $2\PGm2\PGt$ yield. This assumption has been verified with simulated signal samples for two representative mass hypotheses of $\maO=5$ and $10\GeV$. For both masses, the values of ${\mathcal{A}}\epsilon$ for the two decay channels are found to be consistent within 3--5\%. 

\begin{table}[htb]
\centering
\topcaption{
Signal acceptance times selection efficiency ${\mathcal{A}}\epsilon$, defined in the text, and the number of expected signal events after selection in the \SR, computed using simulated signal samples for representative mass hypotheses. The Higgs boson cross sections from \PgPg, VBF, and VH production mechanisms are set to the SM predictions~\cite{deFlorian:2016spz}. The number of expected signal events is computed for a benchmark value of the branching fraction ${\mathcal{B}}(\PH \to \PaO \PaO) {\mathcal{B}}^{2} (\PaO \to \PGt \PGt)=0.05$. The quoted uncertainties in the predictions from simulation include only the statistical component. In data, 7803 events are selected in the \SR.
}
\begin{tabular}{crrrr}
Signal & \multicolumn{2}{c}{\centering ${\mathcal A}\epsilon$ [$10^{-4}$]} & \multicolumn{2}{c}{\centering Expected signal yield} \\
\cline{2-3} \cline{4-5}
$\maO$ [\GeVns{}] & \multicolumn{1}{c}{4\PGt} & \multicolumn{1}{c}{$2\PGm2\PGt$} & \multicolumn{1}{c}{$4\PGt$} & \multicolumn{1}{c}{$2\PGm2\PGt$} \\
\hline
5 & 3.52 $\pm$ 0.10 & 103$\phantom{.}$ $\pm$ 2$\phantom{.1}$ & 134$\phantom{.}$ $\pm$ 4$\phantom{.1}$ & 39.7 $\pm$ 0.4 \\
8 & 2.55 $\pm$ 0.09 & 76.0 $\pm$ 1.0 & 97.2 $\pm$ 3.3 & 23.0 $\pm$ 0.3 \\
12 & 1.37 $\pm$ 0.06 & 35.6 $\pm$ 0.7 & 52.1 $\pm$ 2.4 & 10.1 $\pm$ 0.2 \\
15 & 0.32 $\pm$ 0.03 & 7.5 $\pm$ 0.3 & 12.3 $\pm$ 1.1 & 2.1 $\pm$ 0.1 \\
\end{tabular}
\label{tab:signal_selection}
\end{table}

\section{Final discriminant}
\label{Sec:final_discriminant}
\begin{figure}[ht]
\centering
\includegraphics[width=\textwidth]{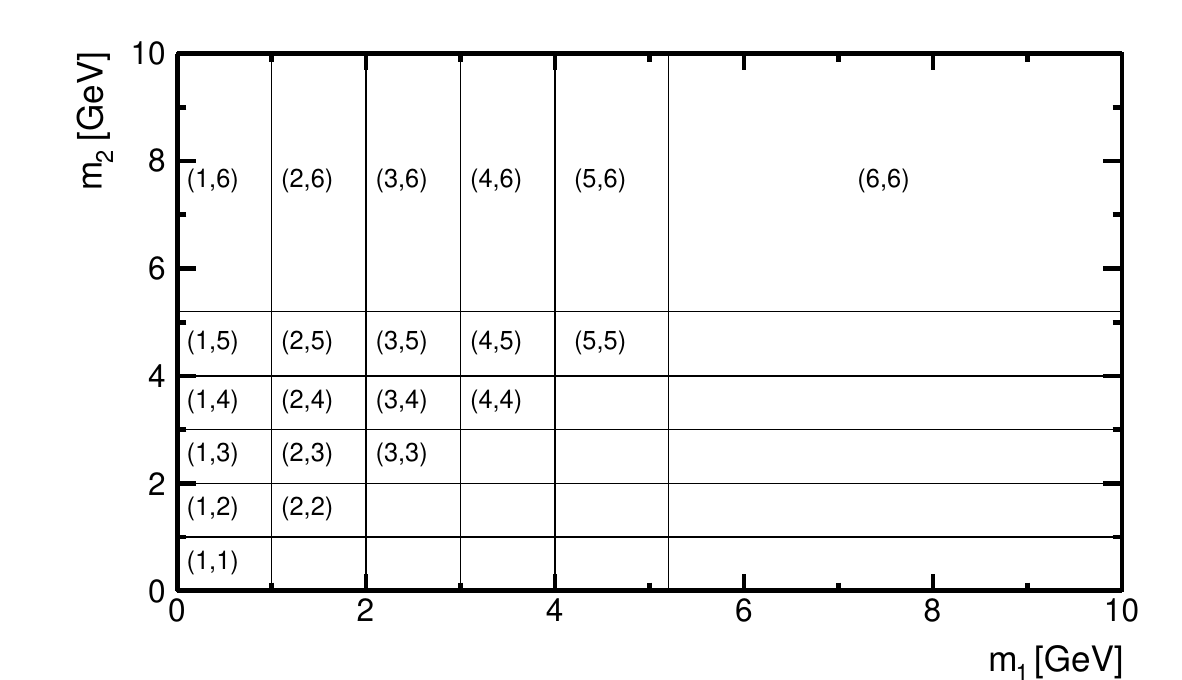}
\caption{Binning of the 2D ($m_1,m_2$) distribution. Each bin is labeled $(i,j)$, where $i$ is the bin index along $m_1$ ($x$ axis) and $j$ is the bin index along  $m_2$ ($y$ axis). Bins $(i,6)$ with $i=1,...,5$ include all events with $m_2>5.2\GeV$, while bin $(6,6)$ contains all events with $m_1$ and $m_2 >5.2\GeV$.}
\label{fig:binning}
\end{figure}

A two-dimensional (2D) distribution of the invariant masses of the muon+track systems constituting the \PaO candidates is chosen as the final discriminant and employed to distinguish between signal and background. This 2D distribution is populated with pairs of muon+track invariant masses ($m_1,m_2$), ordered by their value, $m_2>m_1$. Figure~\ref{fig:binning} illustrates the binning of the 2D distribution. Each bin is labeled $(i,j)$, where $i$ and $j$ denote the bin indices along the $m_1$ and $m_2$ axes, respectively. Since $m_2$ is required to be greater than $m_1$, only the bins $(i,j)$ where $j\ge i$ are filled, resulting in a total of $6 (6+1)/2=21$ independent bins. The bin boundaries along each axis are at 0, 1, 2, 3, 4, and 5.2\GeV. Bins $(i,6)$ with $i=1,\ldots,5$ include all events with $m_2>5.2\GeV$, while bin $(6,6)$ contains all events with $m_1$ and $m_2 >5.2\GeV$. The binning is optimized to achieve the highest sensitivity to the signal for all tested mass hypotheses of \PaO while ensuring reliable and robust estimation of the background, as detailed in Section~\ref{Sec:bkgd}. The expected 95\% \CL upper limit on the signal rate, assuming the presence of only background in data, is employed as the optimization metric. The chosen binning yields, on average, the lowest expected limit for all tested masses of \PaO. At the same time, the statistical uncertainty in the background estimate is kept between 1 and 40\% across all bins, ensuring smooth dependence of the limit on \maO. 

\section{Background modeling}
\label{Sec:bkgd}

As mentioned in Section~\ref{Sec:selection}, QCD multijets constitute the dominant source of background, with small contributions coming from processes such as $\ttbar$, $\PZ$+jets, $\PW$+jets, and diboson production. To model the shape of the ($m_1,m_2$) distribution of the background in the \SR, a binned template is constructed as:
\begin{equation}
f_\text{2D}(i,j) = C(i,j)\, [f_\text{1D}(i)\, f_\text{1D}(j)]^{\text{sym}},
\label{eq:bkgdshape}
\end{equation}
where
\begin{itemize}
\item{$f_\text{2D}(i,j)$ represents the content of bin $(i,j)$ in the ($m_1,m_2$) distribution, normalized to unity.}
\item{$f_\text{1D}(i)$ is the content of bin $i$ in the one-dimensional (1D) muon+track invariant mass distribution, normalized to unity. Each event contributes two entries to this distribution.}
\item{$C(i,j)$ is a symmetric matrix, accounting for possible correlation between $m_1$ and $m_2$. The condition $C(i,j)=1$ for all bins $(i,j)$ would indicate an absence of correlation between $m_1$ and $m_2$. The elements of the matrix $C(i,j)$ are referred to as ``correlation factors'' henceforth. }
\end{itemize}

Equation~(\ref{eq:bkgdshape}) includes a symmetrization operation, denoted by `sym', which is applied to the product of the 1D distributions $f_\text{1D}(i)$ and $f_\text{1D}(j)$ and defined as follows:
\begin{equation}
\left[ f_\text{1D}(i)\, f_\text{1D}(j) \right]^{\text{sym}} = 
\begin{cases} 
2f_\text{1D}(i)\,f_\text{1D}(j), & \text{if } i \neq j, \\
\phantom{2}f_\text{1D}(i)\,f_\text{1D}(i), & \text{if } i = j.
\end{cases}
\label{eq:bkgd1Dsym}
\end{equation}

This symmetrization ensures that the contributions from the nondiagonal bins $(i,j)$ and $(j,i)$ in the Cartesian product $f_\text{1D}(i) f_\text{1D}(j)$ are correctly accounted for in the 2D ($m_1,m_2$) distribution, given the ordering imposed on the muon+track invariant masses. The normalization of the background is left unconstrained prior to the signal extraction.

In order to derive and validate the modeling of $f_\text{1D}(i)$ and $C(i,j)$, multiple control regions (\CR{}s), disjoint from the \SR, are defined based on variations in the isolation criteria applied to one or both muon+track pairs. The isolation criteria are defined by the number of tracks within a cone of $\Delta R=0.5$ around the muon momentum direction. The two muons are labeled as ``first muon'' and ``second muon''. The first muon serves as a ``probe'' used to assess $f_\text{1D}(i)$ as a function of the isolation criteria of the second muon, the ``tag'', as described in Section~\ref{Sec:bkgd_1D} below. In the \CR labeled ``\CRLooseIso'' which is employed to derive the $C(i,j)$, both muons play the role of probes. A summary of the  \CR{}s and \SR, along with the specifications for the first and second muon, is presented in Table~\ref{tab:CR}. Detailed definition of the  \CR{}s is provided in Sections~\ref{Sec:bkgd_1D} and~\ref{Sec:bkgd_corr}. In all  \CR{}s, the expected signal contamination constitutes 0.1--1\% (0.5--6\%) of the background yield across all bins of the $f_\text{1D}(i)$ ($f_\text{2D}(i,j)$) distributions for all tested \PaO boson mass hypotheses, and thus has negligible impact on the background modeling and final result. 

\begin{table}[htb]
\topcaption{Definition of the  \CR{}s used to construct and validate the background model. The last row defines the \SR. The symbols $N_\text{iso}$ and $N_\text{sig}$ denote the number of isolation and signal tracks, respectively, within a cone of $\Delta R=0.5$ around the muon momentum direction. In cases where $N_\text{sig}$ is not mentioned, there is no explicit requirement on the number of signal tracks.}
\label{tab:CR}
\centering
\begin{tabular}{lllp{5cm}}
Region  & First \PGm & Second \PGm  & Purpose\\
\hline
 & & & \\
\CRNtwothree & $N_\text{sig}=1$, $N_\text{iso} = 1$    &
$N_\text{iso}=2,3$  & Determination of $f_\text{1D}(i)$\\
& & & \\
\CRNisoone   & $N_\text{sig}\ge 1$, $N_\text{iso}>1$ & $N_\text{sig}=1$, $N_\text{iso}=1$ & Validation and systematic \\
\CRNisotwothree & $N_\text{sig}\ge 1$, $N_\text{iso}>1$ & $N_\text{iso}=2,3$ & uncertainty estimate of $f_\text{1D}(i)$\\
& & & \\
\CRLooseIso & $N_\text{sig}=1, N_\text{iso}=3,4$ & $N_\text{sig}=1, N_\text{iso}=3,4$ & Determination of $C(i,j)$\\
& & & \\
\SR & {$N_\text{sig}=1$, $N_\text{iso}=1$ } & {$N_\text{sig}=1$, $N_\text{iso}=1$ }  & Signal extraction\\
\end{tabular}
\end{table}

\subsection{Modeling of \texorpdfstring{$f_\text{1D}(i)$}{f[1Di]}}
\label{Sec:bkgd_1D}

The $f_\text{1D}(i)$ distribution is modeled using the \CR labeled ``\CRNtwothree''. This \CR comprises events that pass the same-sign dimuon selection criteria and include only one \PaO candidate, formed by an isolated signal track and a muon (referred to as the first muon). The invariant mass of this first muon and its associated track is used to construct the $f_\text{1D}(i)$ distribution. The second muon in the event must be accompanied by either two or three nearby isolation tracks. Simulations indicate that \CRNtwothree is enriched in QCD events, with less than 5\% of the events coming from non-QCD backgrounds. The modeling of $f_\text{1D}(i)$ is based on the assumption that the kinematic distributions of the muon+track system forming the \PaO candidate are weakly affected by the isolation criteria applied to the second muon, therefore implying that the $f_\text{1D}(i)$ distribution in the \SR is similar to that in \CRNtwothree.

\begin{figure}[hbpt!]
\centering
\includegraphics[width=0.49\textwidth]{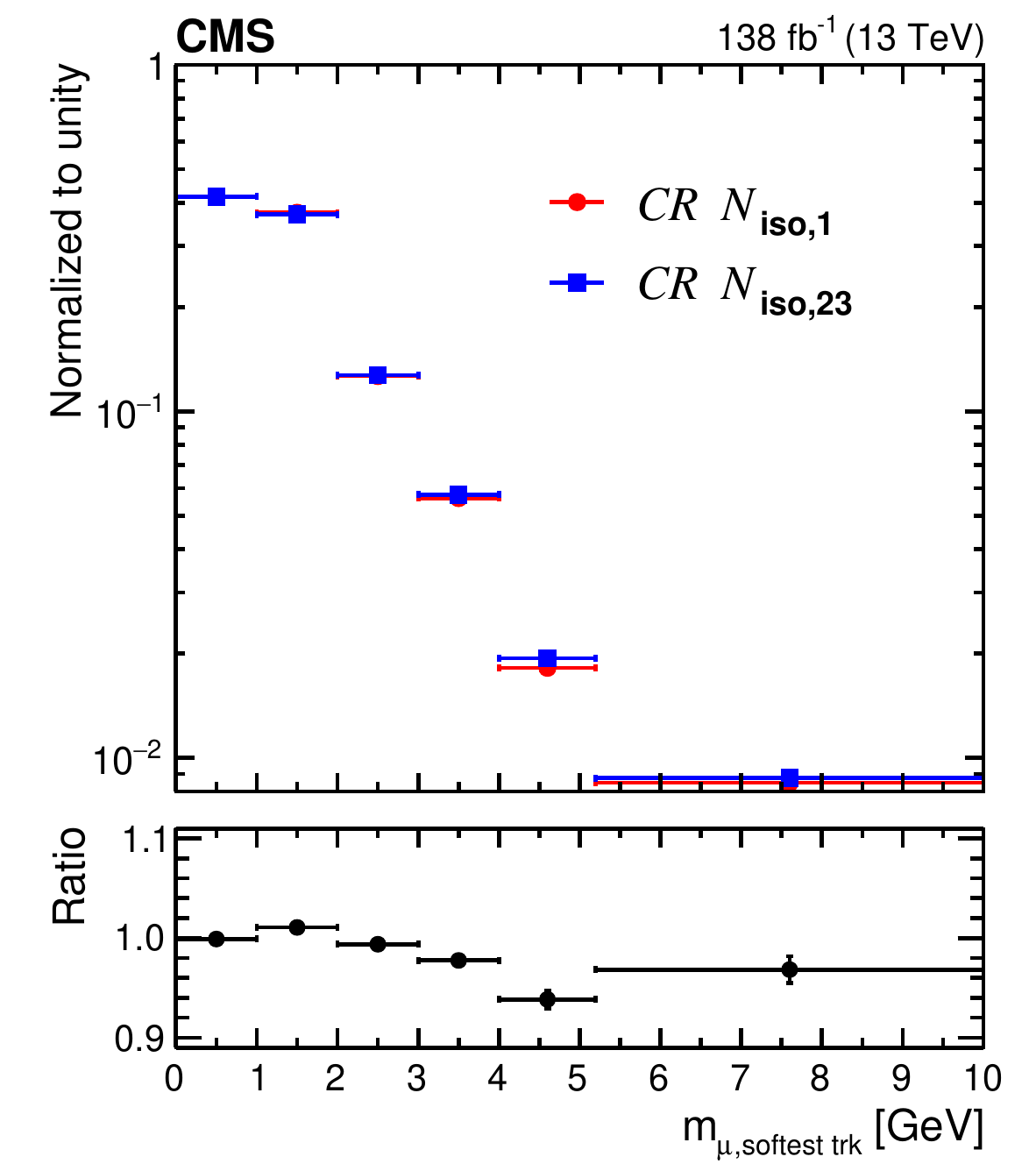}
\includegraphics[width=0.49\textwidth]{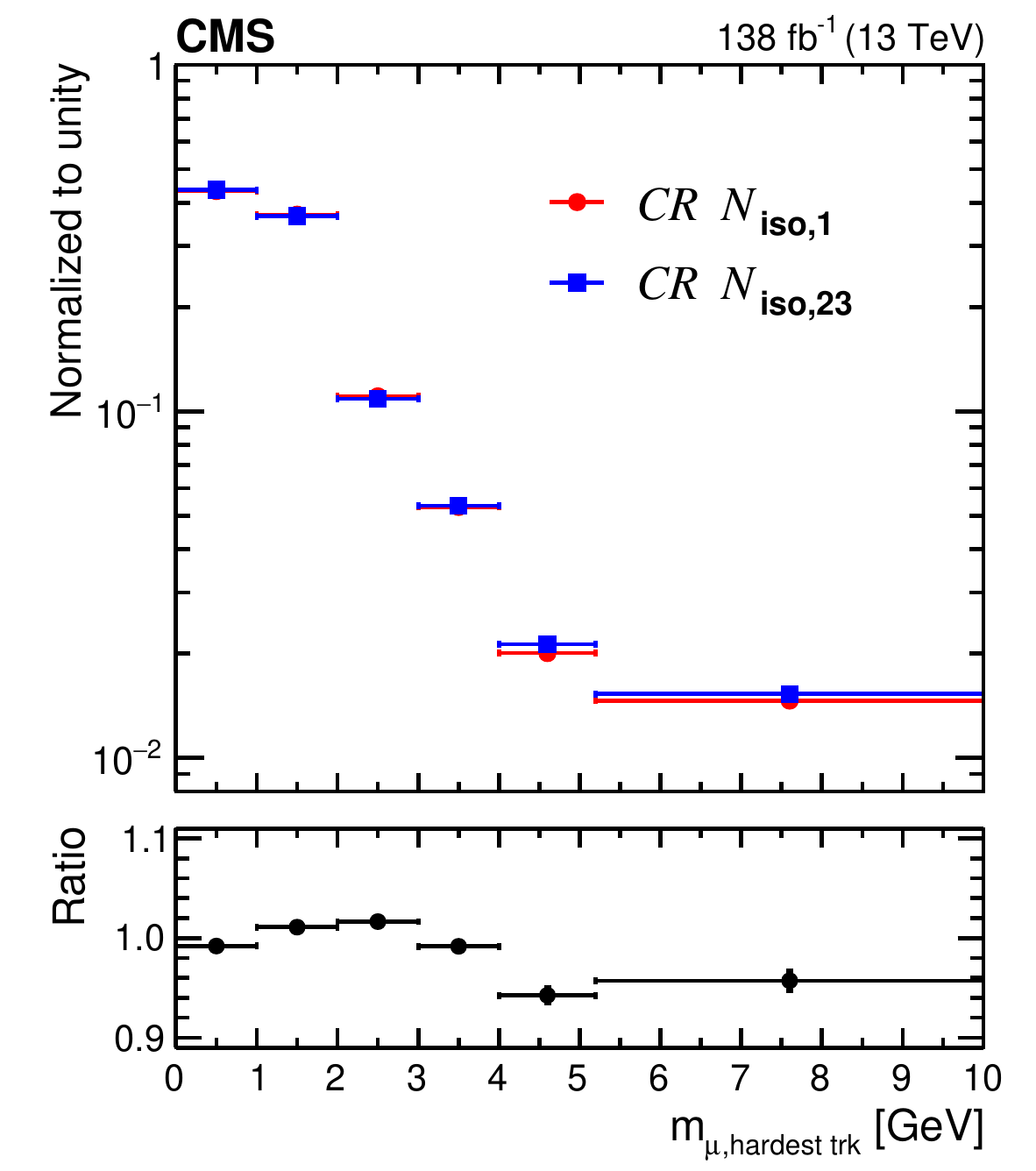}
\caption{
The observed invariant mass distribution, normalized to unity, of the first muon and the softest (left) or hardest (right) accompanying signal track for different isolation requirements imposed on the second muon: one isolation track (\CRNisoone; circles) or two to three isolation tracks (\CRNisotwothree; squares). Vertical bars represent statistical uncertainties, which are smaller than the marker size in most cases and thus imperceptible. The horizontal bars show the bin width. The lower panels show the ratio of the distribution in \CRNisotwothree to that in \CRNisoone. The last bin in both distributions includes all entries with invariant mass of the muon+track system greater than $5.2\GeV$.
}
\label{fig:MassShapesIso}
\end{figure}

A direct test of this assumption is not conclusive because of the limited size of simulated samples. Therefore, the hypothesis is verified using additional  \CR{}s labeled ``\CRNisoone'' and ``\CRNisotwothree''. Events are selected in these  \CR{}s if the first muon has more than one isolation track, and at least one of these isolation tracks also fulfils the criteria for signal tracks. 
Since more than one of these tracks can pass the signal track requirements, two scenarios are evaluated, using either the signal track with the lowest \pt (``softest'') or the highest \pt (``hardest'') to compute the muon+track invariant mass. If only one signal track is found around the first muon, it serves as both the ``softest'' and ``hardest'' track. The two control regions differ in the isolation criteria applied to the second muon: \CRNisoone\ selects events where the second muon has exactly one signal track, matching the isolation condition in the \SR, while \CRNisotwothree\ selects events where it has two or three isolation tracks, similar to \CRNtwothree. The invariant mass distributions of the first muon and its softest or hardest accompanying track are then compared between the two different isolation scenarios of the second muon. 
The results of this study are presented in Fig.~\ref{fig:MassShapesIso}. In both cases, the invariant mass distributions differ in each bin by less than 7\%, suggesting that the invariant mass of the muon+track system forming an \PaO candidate is not highly sensitive to the isolation requirement on the second muon. To address any systematic effects on the modeling of the $f_\text{1D}(i)$ distribution in \CRNtwothree, the observed differences are treated as a shape uncertainty in the normalized $f_\text{1D}(i)$ template.

\begin{figure}[htb!]
\centering
\includegraphics[width=0.75\textwidth]{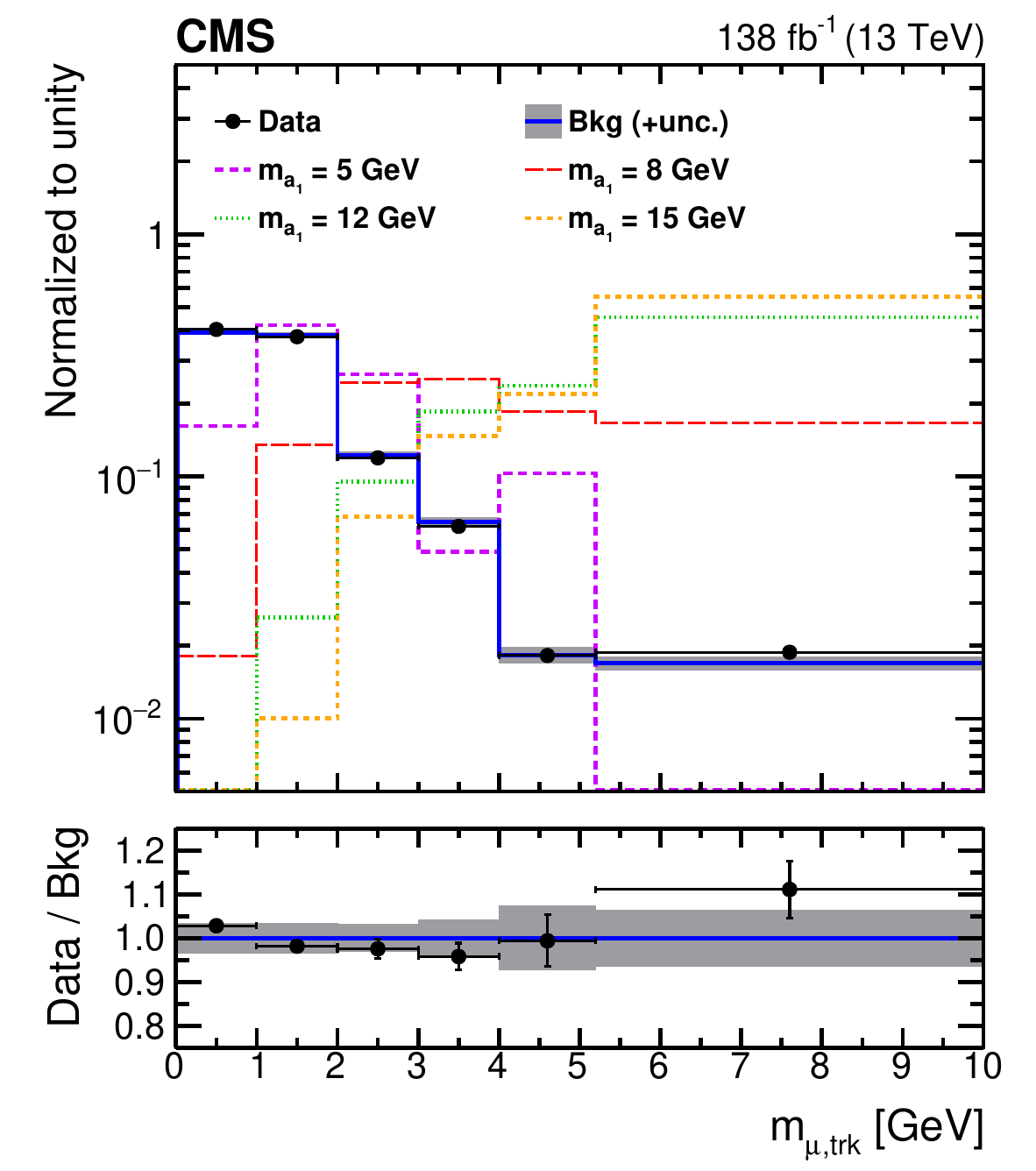}
\caption{
Invariant mass distribution of the muon+track pair, normalized to unity, for events passing the signal selection. Events in data are represented by black points with the vertical bars representing the statistical uncertainty and the horizontal bars the bin width. The expected background distribution derived from \CRNtwothree is shown by the solid blue histogram, with the grey band giving the uncertainty in the background prediction, including systematic and statistical components. Also shown are normalized distributions from signal simulations for four mass hypotheses, $\maO= 5$, 8, 12, and 15\GeV (dashed colored histograms). The lower panel displays the ratio of the data to the expected background.}
\label{fig:MassShapes}
\end{figure}

Figure~\ref{fig:MassShapes} presents the muon+track invariant mass distribution, normalized to unity, for data selected in the \SR and for the background model derived from \CRNtwothree. The data and background distributions are compared to the signal distributions obtained from simulation for four representative mass hypotheses, $\maO= 5$, 8, 12, and 15\GeV. The signal distributions include both the $4\PGt$ and $2\PGm2\PGt$ contributions. Each distribution contains two entries per event, corresponding to the two selected muon+track systems. The invariant mass of the muon+track system demonstrates higher discrimination power between the background and the signal at higher \maO. For lower masses, the signal shape becomes more similar to the background.     

\subsection{Modeling of \texorpdfstring{$C(i,j)$}{C(i,j)}}
\label{Sec:bkgd_corr}

\begin{figure}[hbtp!]
\centering
\includegraphics[width=\textwidth]{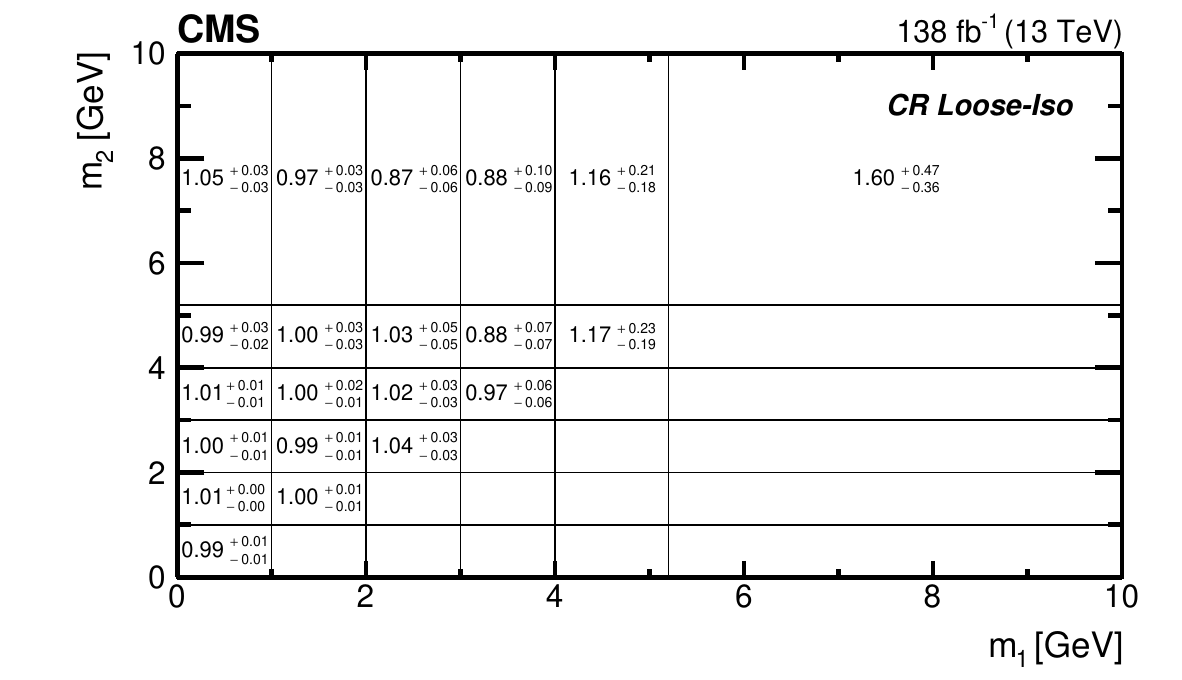}
\caption{
The correlation factors $C(i,j)^{\CR}_{\text{data}}$ with their statistical uncertainties.}
\label{fig:CorrCoeff}
\end{figure}

The correlation factors $C(i,j)$ are determined using a different \CR, labeled ``\CRLooseIso'', which does not overlap with the \SR. This \CR consists of events containing two same-sign muons that meet the identification and kinematic selection criteria detailed in Section~\ref{Sec:selection}. In \CRLooseIso, each muon must have three or four nearby tracks, one of which must be a signal track and the rest are isolation tracks. 
Simulations predict that QCD multijet events comprise approximately 99\% of the events in this \CR. The events in this \CR are used to construct $f_\text{2D}(i,j)$, and the correlation factors $C(i,j)$ are then derived using Eq.~(\ref{eq:bkgdshape}) as:
\begin{equation}
C(i,j) = \frac{f_\text{2D}(i,j)}{[f_\text{1D}(i)\,f_\text{1D}(j)]^{\text{sym}}},
\label{eq:Cij}
\end{equation}
where $f_\text{1D}(i)$ is the 1D normalized distribution with two entries per event. Figure~\ref{fig:CorrCoeff}  shows the correlation factors $C(i,j)^{\CR}_{\text{data}}$ obtained from the data in \CRLooseIso.

\begin{figure}[htb!]
\centering
\includegraphics[width=\textwidth]{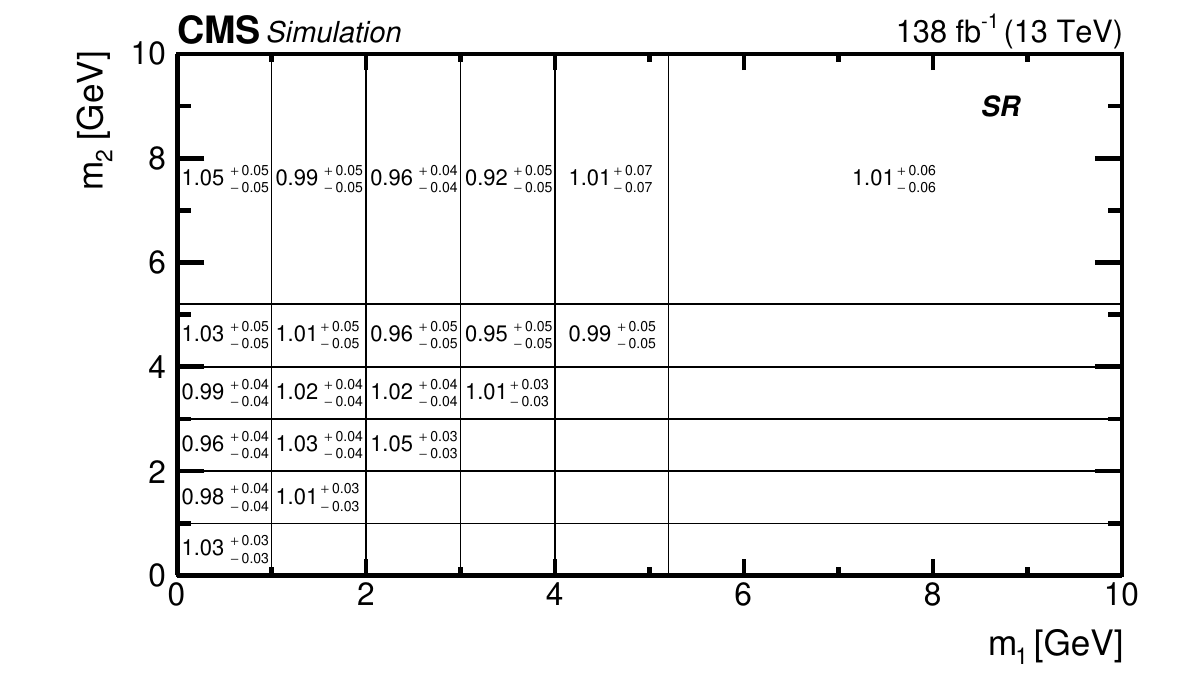}
\includegraphics[width=\textwidth]{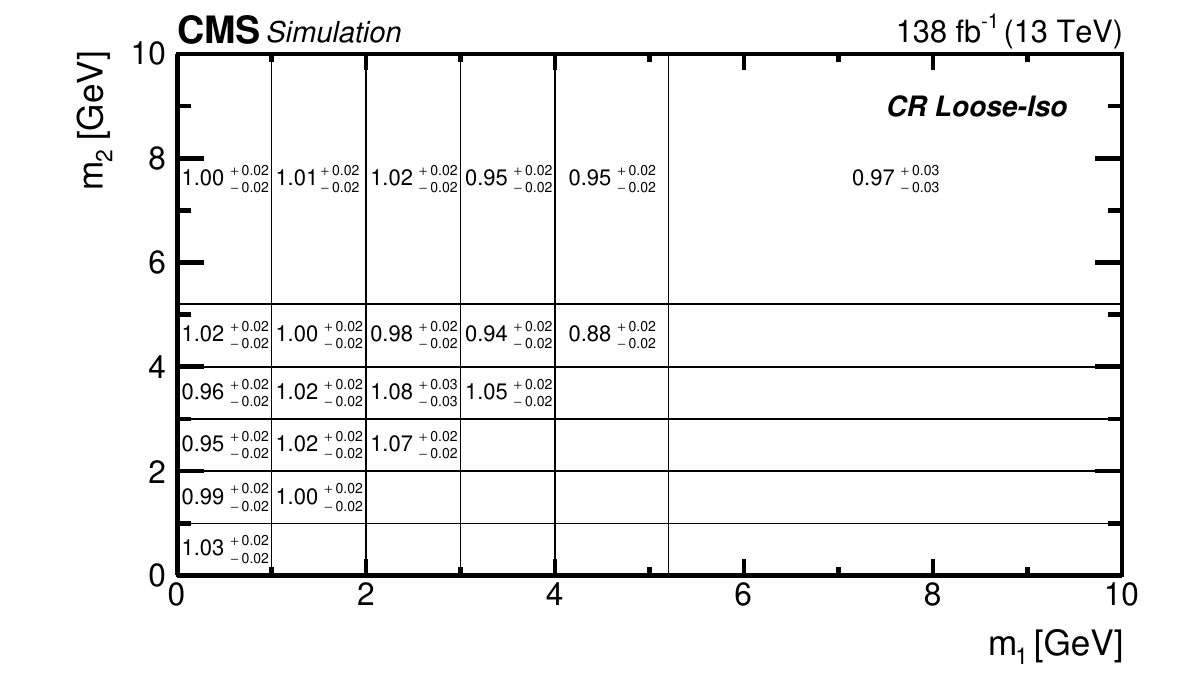}
\caption{
The correlation factors $C(i,j)^{\SR}_{\text{MC}}$ (upper) and  $C(i,j)^{\CR}_{\text{MC}}$ (lower) with their statistical uncertainties.}
\label{fig:CorrCoeff_mc}
\end{figure}

To estimate $C(i,j)$ in data in the \SR, the correlation factors derived with data in \CRLooseIso are corrected for the difference in $C(i,j)$ between the \SR and the \CRLooseIso by comparing samples of simulated background events. The correlation factors estimated from simulation in the \SR, $C(i,j)^{\SR}_{\text{MC}}$, and \CRLooseIso, $C(i,j)^{\CR}_{\text{MC}}$, are shown in Fig.~\ref{fig:CorrCoeff_mc}. As in the \SR, the background in the \CRLooseIso region is dominated by QCD multijet events in which heavy-flavor hadrons decay to muons. Notable differences between the correlation factors derived from data and simulation in this region are observed in bins with large values of $m_1$ or $m_2$, where the contribution from charm- and bottom-hadron decays is highest. These discrepancies are attributed to limitations in the modeling of fragmentation and hadronization effects, and in particular, of the correlated production of heavy-flavor hadrons in separate jets.

The correlation factors in data in the \SR, $C(i,j)^{\SR}_{\text{data}}$, are subsequently computed as
\begin{equation}
C(i,j)^{\SR}_{\text{data}} = C(i,j)^{\CR}_{\text{data}}\,\frac{C(i,j)^{\SR}_{\text{MC}}}{C(i,j)^{\CR}_{\text{MC}}}.
\label{eq:Cij_extrap}
\end{equation}

Finally, the $C(i,j)^{\SR}_{\text{data}}$ values are multiplied by a common scale factor of 1.02 to ensure that the $f_\text{2D}(i,j)$ terms in the \SR add up to unity, according to Eq.~(\ref{eq:bkgdshape}). This scale factor does not affect the final result, as the overall normalization of the background is left unconstrained prior to the signal extraction. 

\section{Signal modeling}
\label{Sec:Signalmodeling}

The signal model template is constructed using simulated samples of $\PH\to\PaO\PaO\to 4\PGt$ and $2\PGm2\PGt$ decays. The analysis measures the signal strength modifier, defined as the product of the measured signal cross section times branching fraction, ${\mathcal{B}}(\PH\to\PaO\PaO){\mathcal{B}}^{2}(\PaO\to \PGt\PGt)$, relative to the value predicted by the SM for the inclusive Higgs boson production cross section. The relative contributions from different Higgs boson production modes are determined by the corresponding cross sections predicted by the SM. The contribution from the $\PH\to\PaO\PaO\to 2\PGm 2\PGt$ decay is computed under the assumption that the partial widths of the $\PaO\to\PGt\PGt$ and $\PaO\to\PGm\PGm$ decays satisfy Eq.~(\ref{eq:width_ratios}).

\begin{figure}[htb]
\centering
\includegraphics[width=0.49\textwidth]{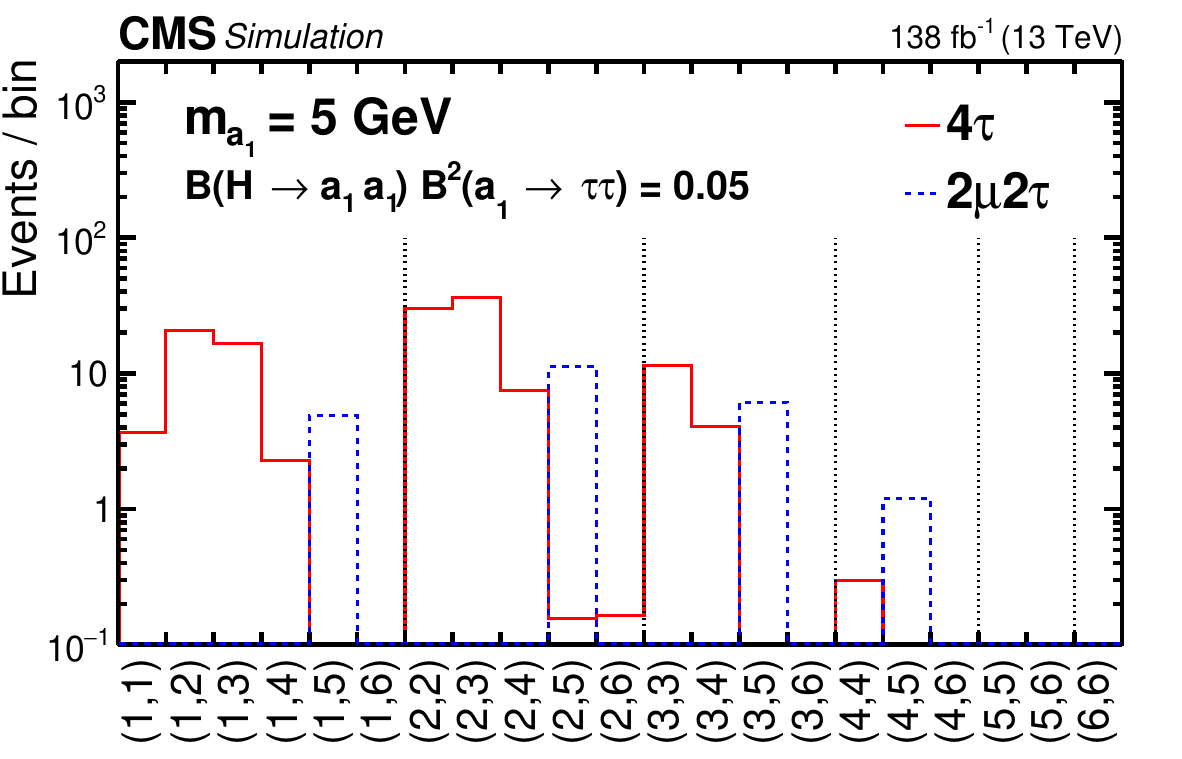}
\includegraphics[width=0.49\textwidth]{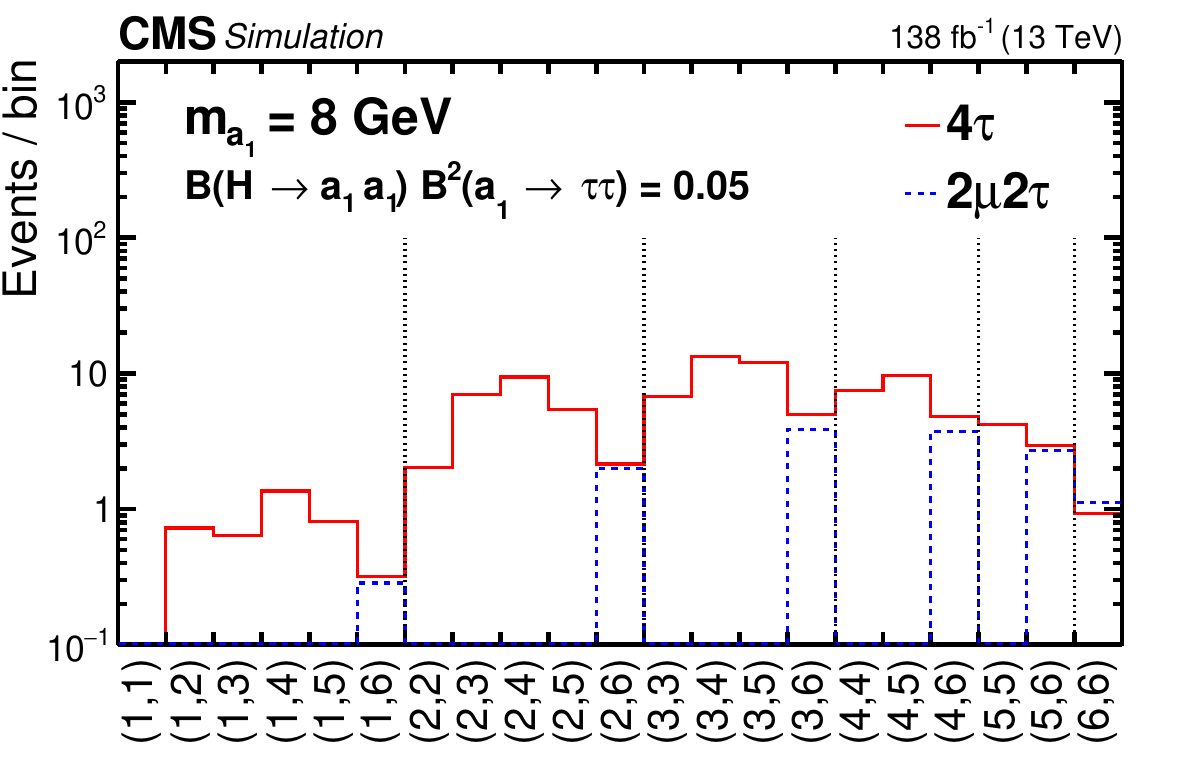}
\includegraphics[width=0.49\textwidth]{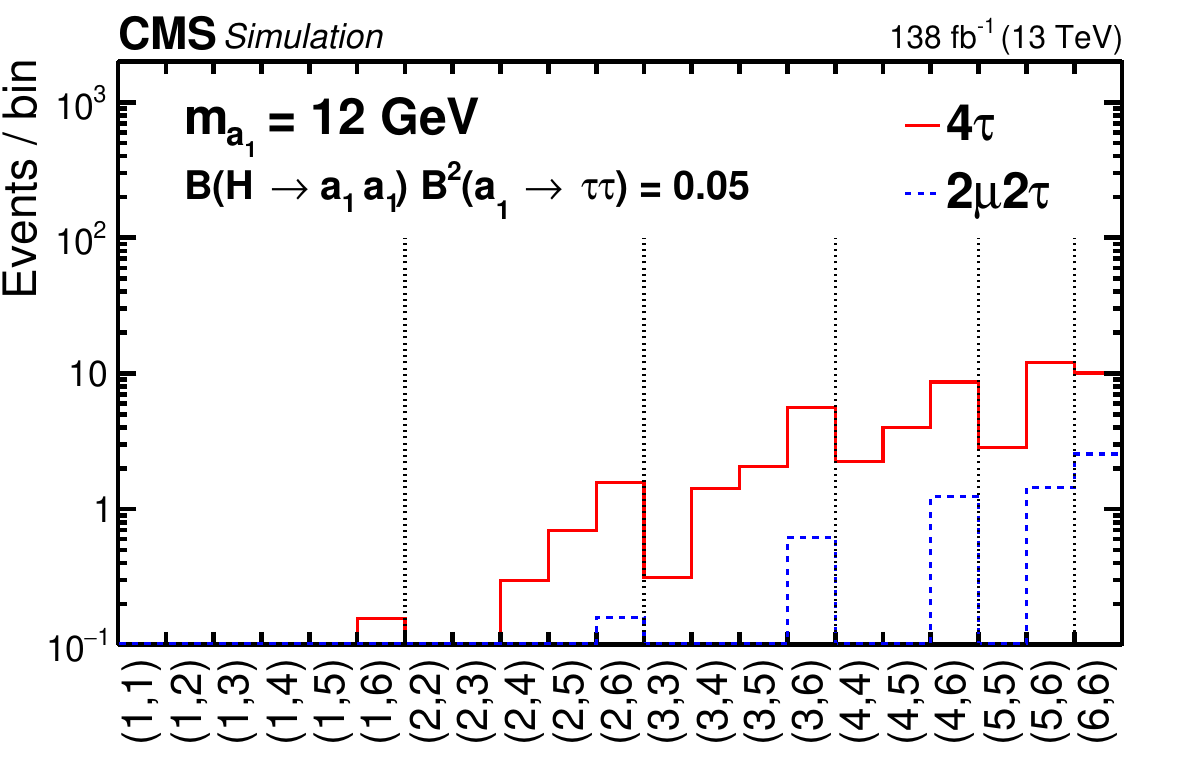}
\includegraphics[width=0.49\textwidth]{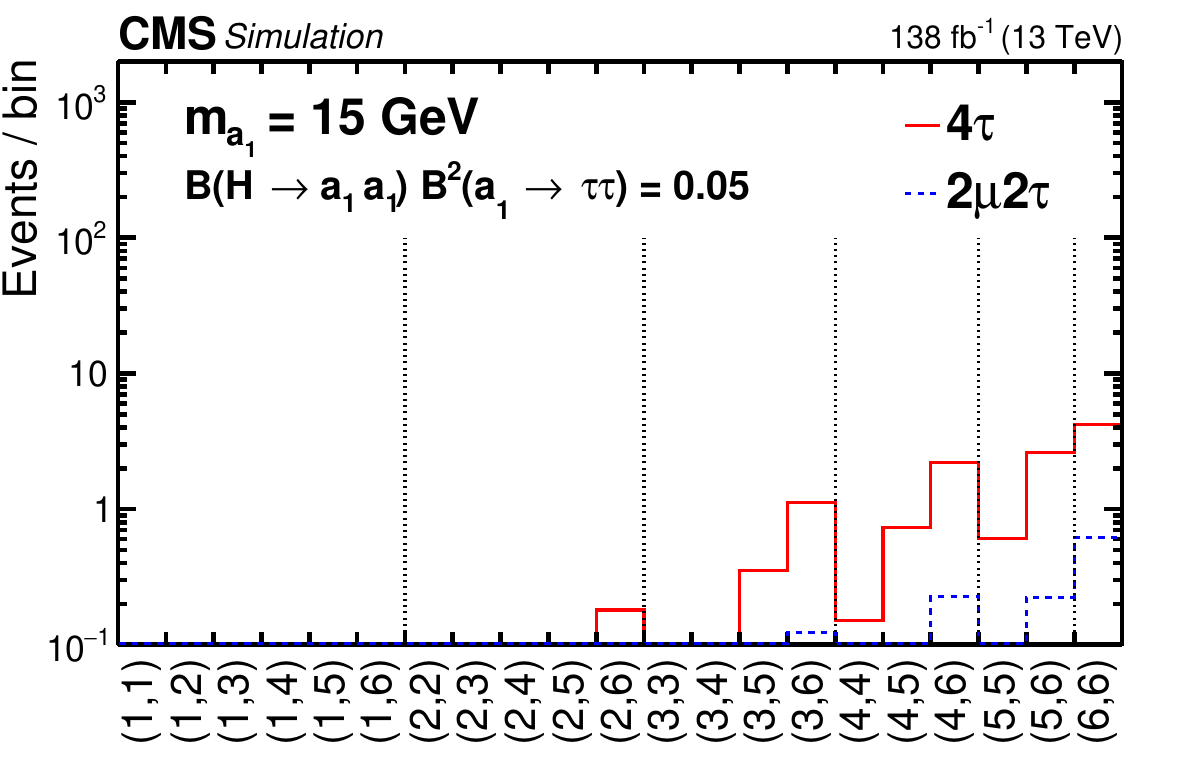}
\caption{
The simulated signal $(m_1, m_2)$ distribution, converted into a 1D array, for \maO values of 5 (upper left), 8 (upper right), 12 (lower left), and 15\GeV (lower right). The contributions of the $\PH\to\PaO\PaO\to 4\PGt$ (red histograms) and $2\PGm2\PGt$ (blue histograms) decays are shown. The distributions are normalized assuming SM Higgs production cross section and ${\mathcal{B}} (\PH\to\PaO\PaO){\mathcal{B}}^{2}(\PaO\to \PGt\PGt)$ = 0.05. The bin notation follows that of Fig.~\ref{fig:binning}.
}
\label{fig:signal_4tau_2mu2tau}
\end{figure}

The muon+track invariant mass distribution in the $\PaO\to\PGm\PGm$ decay channel peaks at the nominal \PaO mass. In contrast, the reconstructed mass of the muon+track system in the $\PaO\to\PGt\PGt$ decay is generally lower and has a larger dispersion due to the presence of undetected neutrinos. As a result, the ($m_1, m_2$) distribution for simulated $2\PGm2\PGt$ events has a considerably different shape than that for $4\PGt$ events. Figure~\ref{fig:signal_4tau_2mu2tau} illustrates the ($m_1, m_2$) distributions, unrolled into a 1D array, for the $4\PGt$ and $2\PGm2\PGt$ simulated events with $\maO=5$, 8, 12, and 15\GeV. The distributions are normalized assuming the SM Higgs boson production rate and ${\mathcal{B}} (\PH\to\PaO\PaO){\mathcal{B}}^{2}(\PaO\to \PGt\PGt) = 0.05$.

\section{Systematic uncertainties}
\label{Sec:Systematics}

Several sources of uncertainty of both statistical and systematic origins are considered in the analysis. A summary of the uncertainties is provided in Table~\ref{tab:uncertainties}. Statistical uncertainties arise from the limited size of the data samples in the  \CR{}s and of the simulated signal samples. These uncertainties are incorporated into the analysis on a bin-by-bin basis using the Barlow--Beeston ``lite'' method~\cite{Barlow:1993dm}. Various systematic uncertainties are included,  classified into two categories: background-related and signal-related.

\begin{table}[th]
    \centering
    \topcaption{Summary of systematic uncertainties affecting the estimation of signal and background. The terms ISR and FSR refer to initial- and final-state radiation, and the symbols $\mu_\text{R}$ and $\mu_\text{F}$ denote the renormalization and factorization scales, respectively. The impact of shape-altering and bin-by-bin uncertainties is quoted in terms of relative variations of yields across all bins of the modeled $(m_1,m_2)$ distributions. For the normalization (norm.)~uncertainties, the impact on the overall estimated yield is reported. The last column indicates how a given uncertainty is correlated across the data-taking years. Bin-by-bin statistical uncertainties for simulated signal samples are quoted for the most populated bins containing 80\% of the total yield of selected signal events.}
    \label{tab:uncertainties}
    \begin{tabular}{lcrc}
     Source                                         & Type & Value & Correlated \\
     \hline 
     & & & \\
     \multicolumn{4}{c}{Uncertainties affecting background estimate} \\
     Modeling of $f_\text{1D}(i)$ in  \CR{}s            & shape & 2--7\%     & no \\
     ISR/FSR scales                                 & shape & 1--2\%      & yes \\
     Non-QCD contribution                           & shape &  6\%        & yes \\     
     Limited size of data samples in  \CR{}s            & bin-by-bin& 1--40\% & no \\
     & & & \\
    \multicolumn{4}{c}{Uncertainties affecting signal estimate} \\
     Integrated luminosity                                     & norm. & $<$3\%    & partially \\ 
     Muon ID and trigger efficiency                 & norm. &  3\%      & yes \\
     Track ID/isolation efficiency                  & shape & 10--20\%  & no \\
     Prefiring weights                              & norm. & $<$3\%    & no \\
     $\PQb$ tagging                                & norm. & $<$0.5\%  & partially \\ 
     Limited size of simulated samples              & bin-by-bin & 1--2\% & no \\ 
     $\mu_\text{R}$ and $\mu_\text{F}$ scales (${\mathcal{A}}\epsilon$) & norm. & 1--3\% & yes \\
     $\mu_\text{R}$ and $\mu_\text{F}$ scales (cross sections)    & norm. & 1--5\%    & yes \\
     PDF (${\mathcal{A}}\epsilon$)                & norm. & 1--2\%    & yes \\
     PDF (cross sections)                           & norm. & 1--3\%    & yes \\
    \end{tabular}
\end{table}

\subsection{Uncertainties related to background}

The shape of the background in the ($m_1,m_2$) distribution is modeled according to Eq.~(\ref{eq:bkgdshape}). This distribution is affected by the shape uncertainty in $f_\text{1D}(i)$, described in Section~\ref{Sec:bkgd_1D}, which accounts for potential biases introduced by estimating the 1D distribution in \CRNtwothree. The uncertainty, derived by comparing $f_\text{1D}(i)$ between \CRNisoone and \CRNisotwothree, is found to vary from 2--7\% across individual bins of $f_\text{1D}(i)$. These variations are propagated to $f_\text{2D}(i,j)$ in Eq.~(\ref{eq:bkgdshape}) via six nuisance parameters, one per bin of the $f_\text{1D}(i)$ template. 

The background shape is further impacted by uncertainties related to the extrapolation of $C(i,j)$ from \CRLooseIso to the \SR using simulated background samples. Variations in the modeling of initial- and final-state radiation (ISR and FSR) during parton showering can impact the reconstructed muon+track invariant masses and their correlations in events where an $\PaO$ candidate is mimicked by a hadronic jet, thus leading to potential deviations in $C(i,j)$. To account for this, the ISR and FSR parton shower scales are varied independently up and down by a factor of 2. These variations change the estimate of $C(i,j)$ in \CRLooseIso and the \SR by 0.5--3\% depending on $(i,j)$. Uncertainties in the estimate of $C(i,j)$ also arise from limited understanding of contributions from non-QCD events. The non-QCD background fraction is varied by $\pm 50$\%, leading to variations between 0.2 and 5\% across $(i,j)$. Shape-altering uncertainties in the estimation of $C(i,j)^{\CR}_{\text{data}}$ are incorporated by varying the correlation factors $C(i,j)^{\SR}_{\text{MC}}$ and $C(i,j)^{\CR}_{\text{MC}}$ based on the aforementioned systematic shifts. The associated shape uncertainties are determined by comparing the correlation factors derived from simulated events in the \SR and \CRLooseIso after applying these systematic variations. Overall, the systematic uncertainties associated with ISR, FSR, and non-QCD contributions induce variations of up to 1, 2, and 6\%, respectively, in the background yields for individual bins. Other uncertainties related to the simulation of background samples have a negligible impact on $C(i,j)$ and affect the final results only marginally.

\subsection{Uncertainties related to signal}

The integrated luminosities for the data-taking years 2016, 2017, and 2018 have individual uncertainties in the range 1.2--2.5\%~\cite{CMS-LUM-17-003,CMS-PAS-LUM-17-004,CMS-PAS-LUM-18-002}, while the overall uncertainty for the 2016--2018 period is 1.6\%.

Uncertainties in muon identification and trigger efficiencies, as determined using a ``tag-and-probe'' method~\cite{CMS:2011aa}, are estimated to be 1.5\% per muon. The efficiencies of track selection and muon+track isolation are evaluated in a study of $\PZ\to\PGt\PGt$ events where one tau lepton decays to a muon while the other decays into an isolated track in a 1-prong decay. This track meets the same selection criteria as in the nominal analysis. In this study, scale factors, accounting for differences in efficiency between data and simulation, are derived and applied to the simulated samples. The uncertainties in scale factors affect the shape of the signal estimate and alter the overall signal yield by 10--20\%.

During the 2016--2017 data-taking periods, a timing shift in the ECAL L1 trigger inputs from the forward endcap region ($\abs{\eta} > 2.4$) caused inefficiencies by incorrectly associating events with the previous bunch crossing~\cite{CMS:2020cmk}. A correction for this effect, determined using an unbiased data sample, is applied to the simulation, accompanied by normalization-altering uncertainties ranging between 0.1 and 2.8\%, depending on the \PaO boson mass and signal sample.

The uncertainties in measuring the \PQb tagging efficiency are applied separately to heavy-flavor and light-flavor jets in the simulated samples as described in Ref.~\cite{CMS-DP-2023-005}. These uncertainties are divided into components specific to the data-taking period and components correlated across periods. The \PQb tagging uncertainties lead to variations in the yield between 0.2 and 0.5\%.

Theoretical uncertainties impact the kinematic distributions of the Higgs boson, particularly its \pt spectrum, thereby affecting the signal acceptance. The uncertainty due to missing higher-order corrections in the \PgPg process is estimated using the HqT program by varying independently the renormalization ($\mu_\text{R}$) and factorization ($\mu_\text{F}$) scales by factors of 0.5 and 2. The \pt-dependent $K$-factors are recomputed according to these variations and applied to the simulated signal samples. The resulting effect on the signal acceptance varies between 2.5 and 3\%, depending on \maO. Similarly, uncertainties in the signal acceptance for the VBF and VH processes are computed, with impacts ranging from 1 to 3\%, according to the process and \maO.

The HqT program is also used to evaluate the uncertainties arising from the choice of PDFs for the \PgPg process. Nominal $K$-factors for the \pt spectrum of the Higgs boson are computed using the NNPDF3.1 PDF set~\cite{Ball:2014uwa}. Variations within the uncertainties in the NNPDF3.1 PDFs alter the signal acceptance by approximately 1\%. The impact of the PDF uncertainties on the acceptance of the VBF and VH processes is estimated in a similar way, resulting in a 2\% uncertainty.

Systematic uncertainties in the theoretical predictions for the Higgs boson production cross section are driven by variations of the $\mu_\text{F}$ and $\mu_\text{R}$ scales and the PDF uncertainties. Uncertainties related to scale variations range from 1 to 5\%, depending on the Higgs boson production mode, whereas the uncertainties related to PDF vary between 1 and 3\%.

Bin-by-bin statistical uncertainties in $C(i,j)$, related to the limited size of the data sample in \CRLooseIso, constitute the dominant uncertainty across all probed mass hypotheses. Additionally, shape uncertainties related to the modeling of $f_\text{1D}(i)$ have a substantial effect. For higher \maO, the uncertainty associated with the track selection and muon+track isolation efficiency also becomes significant.

\section{Results}
\label{Sec:Results}
The signal is extracted with a binned maximum-likelihood fit applied to the 2D ($m_1,m_2$) distribution, using the CMS statistical analysis tool \textsc{Combine}~\cite{CMS:2024onh}, based on the \textsc{RooFit}~\cite{Verkerke:2003ir} and \textsc{RooStats}~\cite{Moneta:2010pm} frameworks. 
The normalizations of both the signal and background are allowed to float freely in the fit. Systematic uncertainties affecting the normalization of the signal templates are incorporated via nuisance parameters with log-normal distributions. Shape-altering systematic uncertainties are modeled by nuisance parameters whose variations cause continuous morphing of the signal or background template shapes, as discussed in Section 4.2.1 of Ref.~\cite{CMS:2024onh}, and are assigned Gaussian prior probability density functions. For each probed \maO, the ($m_1,m_2$) distribution is fitted with the sum of five templates: one for the background model and four for the signal. The signal templates correspond to the \PgPg, VBF, and VH production modes with $\PH\to\PaO\PaO\to 4\PGt$, and the inclusive Higgs boson production followed by $\PH\to\PaO\PaO\to 2\PGm 2\PGt$. The normalization of the signal templates is scaled by a common signal strength modifier. The yields of the $4\PGt$ and $2\PGm2\PGt$ signals are related according to Eqs.~(\ref{eq:width_ratios}) and~(\ref{eq:br_ratios}). No significant excess of events over the SM background prediction is observed. The compatibility of the observed ($m_1,m_2$) distribution with the background-only model is quantified with a goodness-of-fit test based on the saturated model for the test statistics~\cite{parametric_stat_inf, Cousins:2018tiz}, yielding a $p$-value of 0.45. 

\begin{figure}[htb!]
\centering
\includegraphics[width=\textwidth]{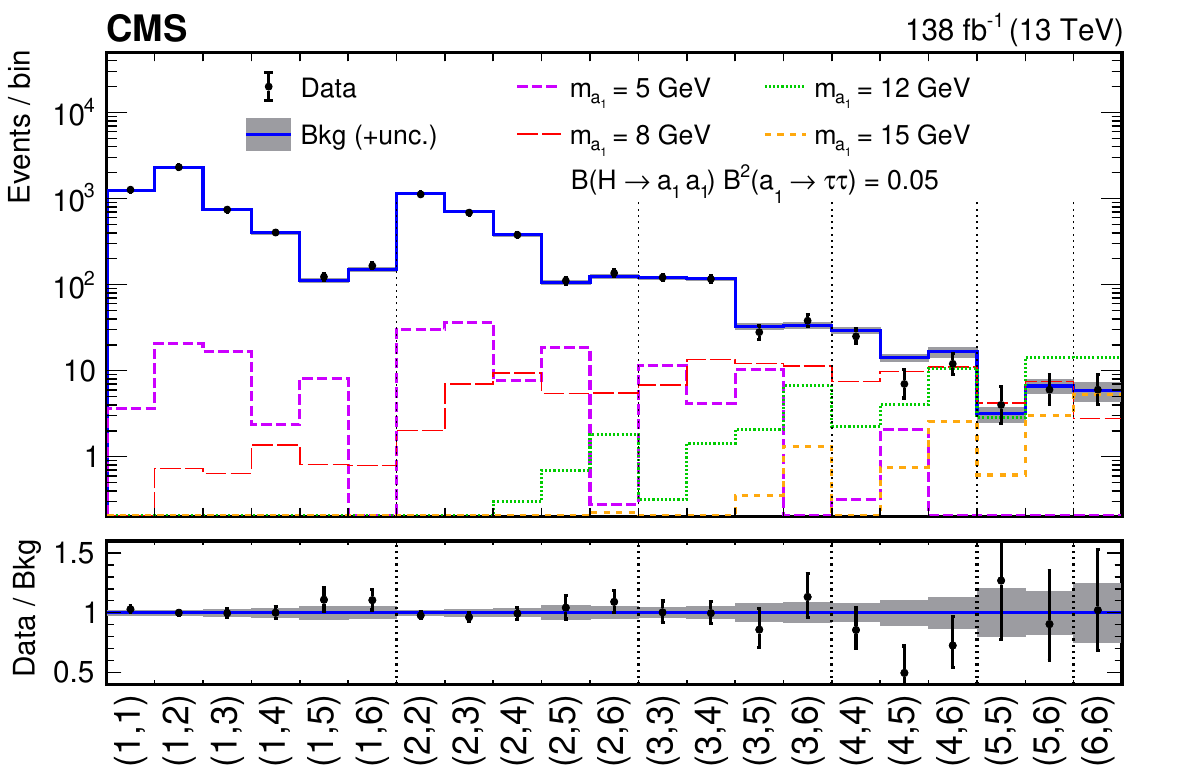}
\caption{
The unrolled ($m_1,m_2$) distribution used to extract the signal. The observed number of events in data is represented by the points, with the vertical bars giving the statistical uncertainty. The background is shown as the blue histogram, with its uncertainty depicted by the shaded grey band. The normalization for the background is obtained by fitting the observed data under the background-only hypothesis. Signal expectations for the $4 \PGt$ and $2\PGm2\PGt$ final states are shown as dashed histograms for the mass hypotheses $\maO= 5$, 8, 12, and 15\GeV. The relative normalization of the $4 \PGt$ and $2\PGm2\PGt$ final states is given by Eq.~(\ref{eq:width_ratios}). The signal normalization is computed assuming that the Higgs boson is produced in $\Pp\Pp$ collisions with a rate predicted by the SM and decays into the $\PaO \PaO \to 4\PGt$ final state with a branching fraction of 5\%. The lower plot shows the ratio of the observed data events to the expected background yield in each bin of the ($m_1,m_2$) distribution.
}
\label{fig:AnalysisBins}
\end{figure}

Figure~\ref{fig:AnalysisBins} displays the unrolled ($m_1,m_2$) distribution, where the notation for the bins follows that of Fig.~\ref{fig:binning}. For illustrative purposes, the background distribution is normalized by fitting the observed data under the background-only hypothesis. Expectations for the signal for $\maO= 5$, 8, 12, and 15\GeV are also shown. The signal normalization is calculated assuming the SM prediction for the cross sections of the \PgPg, VBF, and VH processes and a branching fraction of 5\% for the $\PH \to \PaO \PaO \to 4\PGt$ decays.  

\begin{figure}[htb]
\centering
\includegraphics[width=0.75\textwidth]{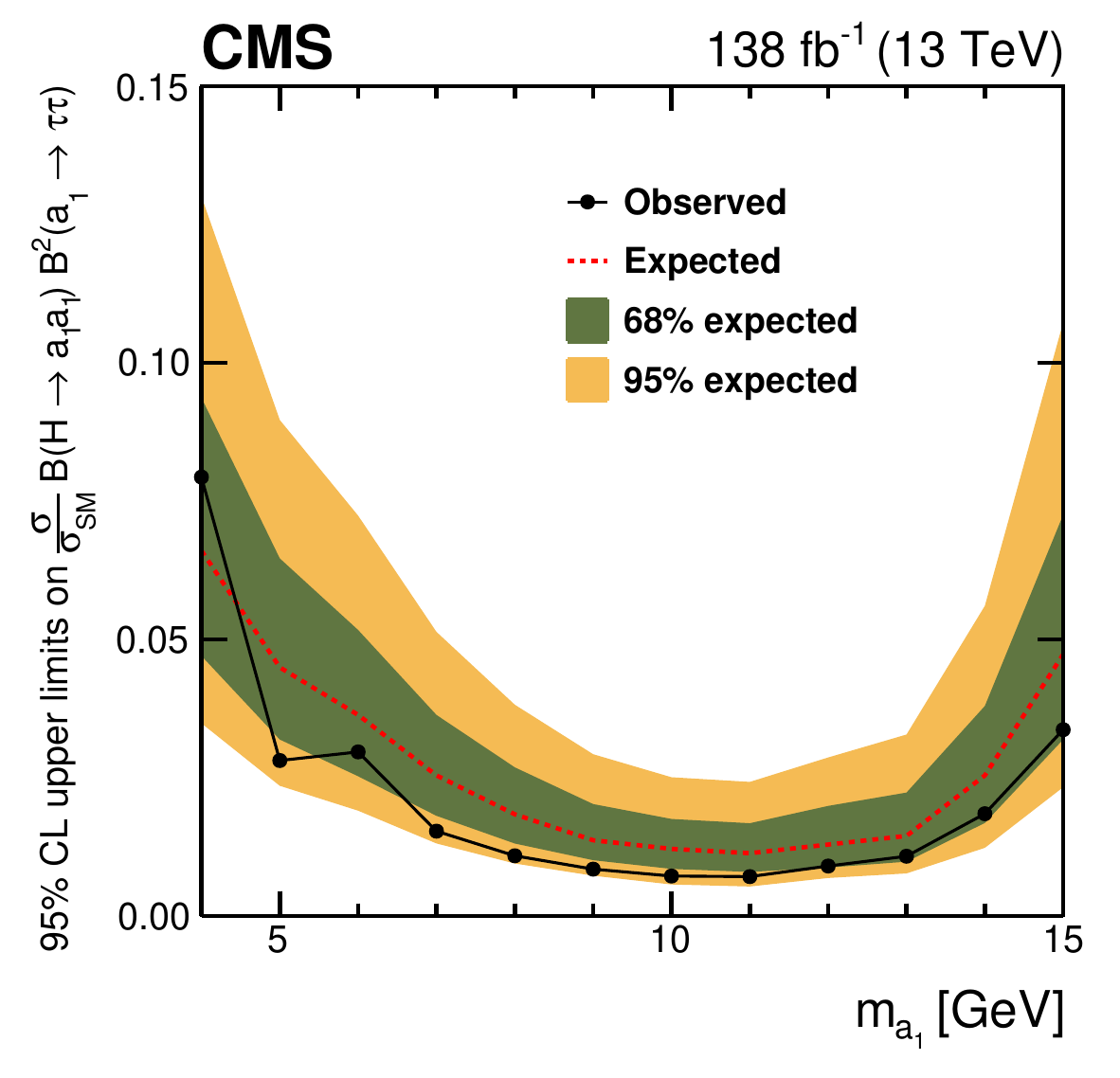}
\caption{
The observed (points) and expected (red line) 95\% \CL upper limits on the product of the signal cross section and the branching fractions $\sigma (\Pp\Pp \to \PH+\PX) {\mathcal{B}} (\PH \to \PaO \PaO) {\mathcal{B}}^{2} (\PaO \to \PGt \PGt)$, relative to the inclusive Higgs boson production cross section $\sigma_\text{SM}$ predicted in the SM. The green and yellow bands indicate the regions containing 68 and 95\% of the expected limit ranges under the background-only hypothesis.
\label{fig:limits}
}
\end{figure}

The results of the analysis are used to set upper limits at 95\% \CL on the product of the cross section and branching fractions, $\sigma (\Pp\Pp \to \PH+\PX) {\mathcal{B}} (\PH \to \PaO \PaO) {\mathcal{B}}^{2} (\PaO \to \PGt \PGt)$, relative to the inclusive SM Higgs boson production cross section, $\sigma_\text{SM}$. The $\mathrm{CL_{s}}$ criterion~\cite{Junk:1999kv, Read:2002hq} is used to set the upper limits, using the asymptotic approximation~\cite{Cowan:2010js}. The test statistic employed in the statistical inference is the profile likelihood ratio modified for upper limits~\cite{CMS-NOTE-2011-005}. Figure~\ref{fig:limits} shows the obtained observed and expected upper limits. The observed limits vary from 0.007 at $\maO=11\GeV$ to 0.079 at $\maO=4\GeV$.  The expected upper limits in the absence of signal range between 0.011 at $\maO=11\GeV$ to 0.066 at $\maO=4\GeV$. The observed upper limits are compatible with the expected limits within two standard deviations over the entire \maO range. 

An assessment of the pre-fit agreement in the shape of the ($m_1,m_2$) distribution between the data and the background model, after normalizing the background distribution to the total number of events observed in the \SR, reveals deficits in the observed yields primarily along the diagonal and near-diagonal bins. These bins contain the bulk of the $\PH\to\PaO\PaO$ signal events, as expected for the decay of the Higgs boson into two identical bosons. The most pronounced deviations occur in bins (2,2), (2,3), (4,4), (4,5), and (4,6). This pattern results in observed upper limits that are stronger than the expected across a broad range of \maO. The largest discrepancy between the observed and expected limits amounts to 1.7 standard deviations, at $\maO=8\GeV$. Note that since the pre-fit background distribution is normalized to the total event yield in data, the deficits in the aforementioned bins are accompanied by excesses in other bins, notably in the off-diagonal bins (1,5), (1,6), and (2,6), and diagonal bin (1,1). It should also be emphasized at this point that the background distribution shown in Fig.~\ref{fig:AnalysisBins} is obtained after performing the fit to data under the background-only hypothesis, in which the differences between the observed and expected yields in individual $(m_1, m_2)$ bins are reduced by variations of the nuisance parameters included in the statistical model.

The decrease in sensitivity observed at lower \maO values is caused by the increase in background towards smaller muon+track invariant mass, as illustrated in Figs.~\ref{fig:MassShapes} and~\ref{fig:AnalysisBins}. As \maO increases, the average angular separation between the decay products of the \PaO boson increases. Consequently, the efficiency of the signal selection decreases due to the requirement that the muon and the track from the $\PaO \to \PGt_{\PGm}\PGt_{\text{1-prong}}$ or $\PaO\to\PGm\PGm$ decay must be within a cone of $\Delta R=0.5$. This explains the reduced sensitivity at higher values of \maO.

The inclusion of the $\PH \to \PaO\PaO\to 2\PGm 2\PGt$ channel in the signal model consistently improves both the expected and observed limits across all mass points, with improvements ranging from 25 to 35\% in the expected limits and up to 30\% in the observed limits. The sizable improvement of sensitivity with the inclusion of the $2\PGm2\PGt$ channel is driven by two factors. Firstly, ${\mathcal{A}}\epsilon$ for the $2\PGm2\PGt$ channel is higher compared to that for $4\PGt$ because of the substantially harder \pt spectrum of muons in $\PaO\to\PGm\PGm$ decays than in $\PaO \to \PGt_{\PGm}\PGt_{\text{1-prong}}$ decays. Secondly, the $\PaO\to\PGm\PGm$ decay produces resonant structures in the $m_1$ distribution, making the discrimination between signal and background in the $2\PGm2\PGt$ channel more efficient than in the $4\PGt$ channel.  

\begin{figure}[t]
\centering
\includegraphics[width=0.49\textwidth]{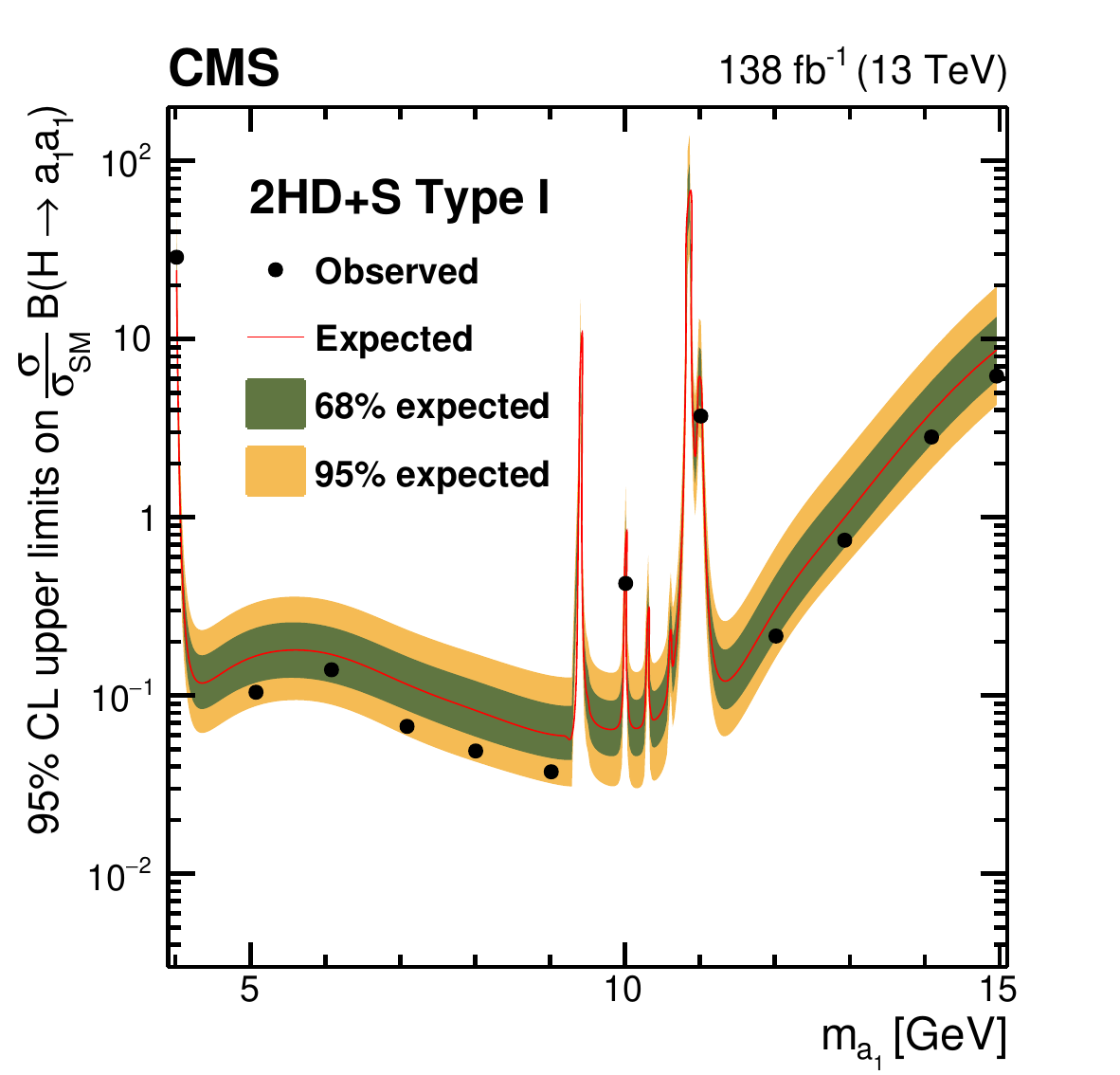}
\includegraphics[width=0.49\textwidth]{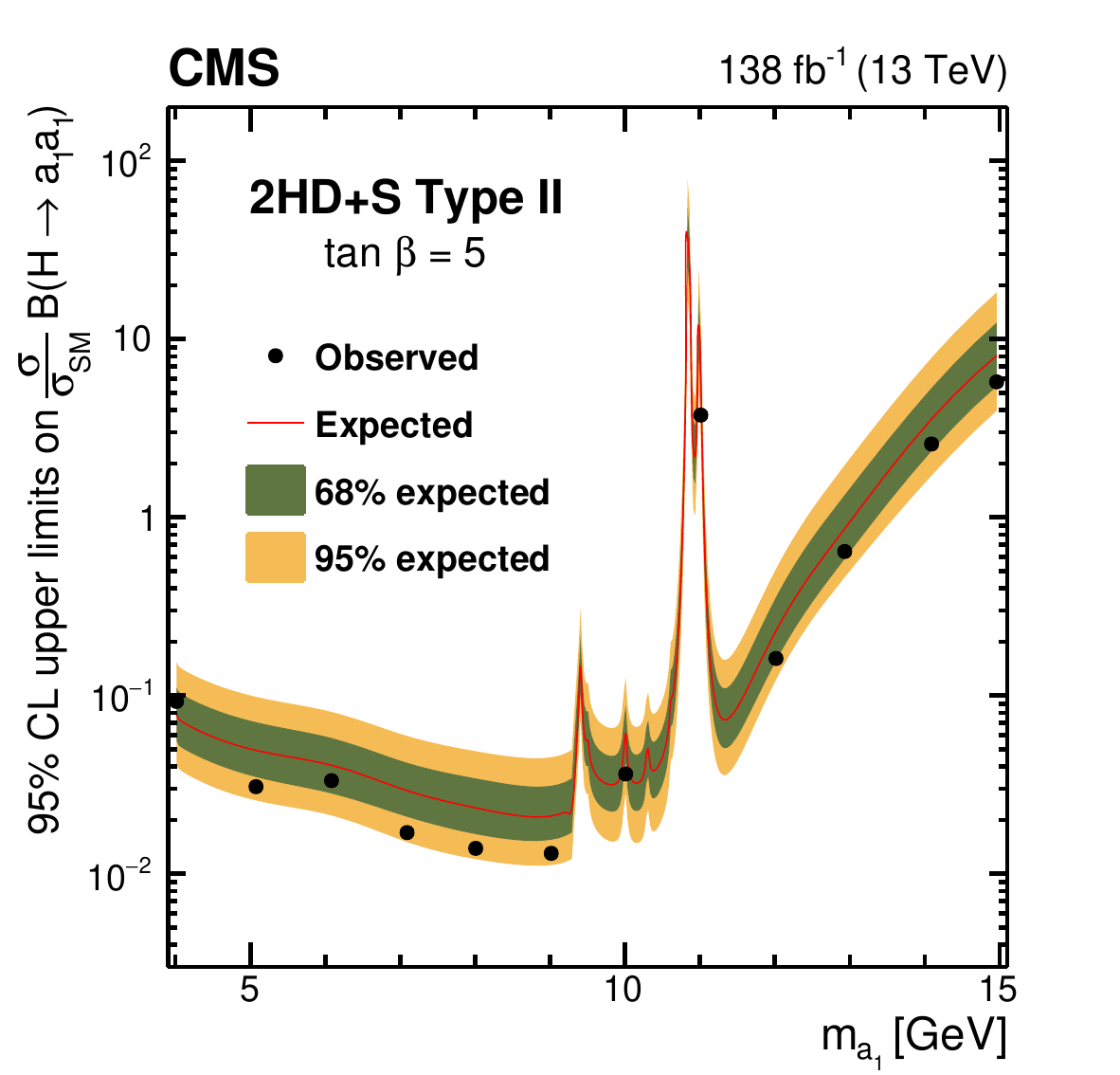}
\includegraphics[width=0.49\textwidth]{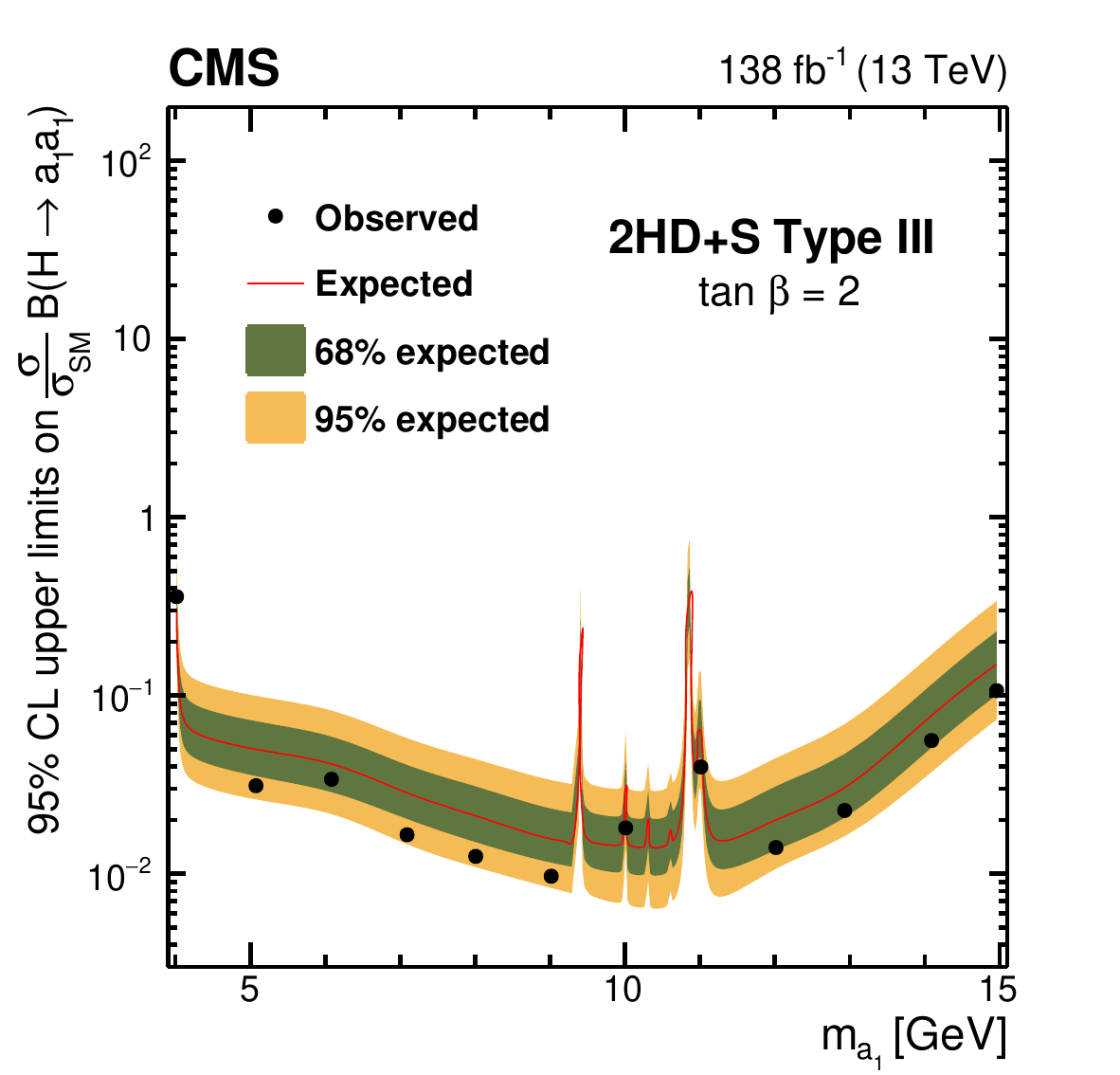}
\includegraphics[width=0.49\textwidth]{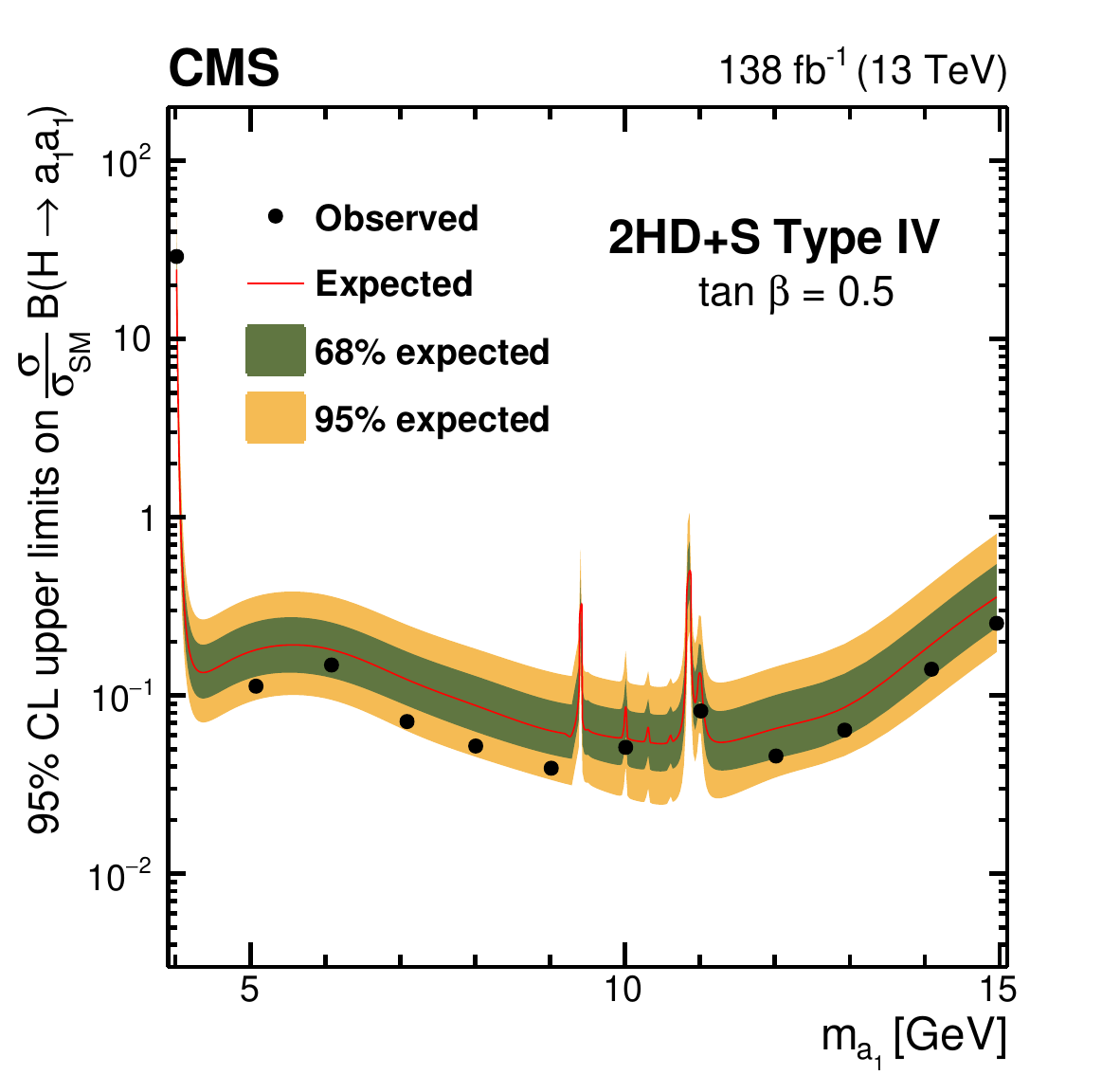}
\caption{
The observed (points) and expected (red line) 95\% \CL upper limits on $\sigma (\Pp\Pp \to \PH+\PX) {\mathcal{B}} (\PH \to \PaO \PaO)$, relative to $\sigma_\text{SM}$, as a function of \maO for different 2HD+S models at benchmark $\tan\beta$ values: Type~I ($\tan\beta$ independent; upper left), Type~II ($\tan\beta = 5$; upper right), Type~III ($\tan\beta = 2$; lower left), and Type~IV ($\tan\beta = 0.5$; lower right).
\label{fig:modeldependentlimits2D}
}
\end{figure}

The results of the search are also interpreted in the context of the 2HD+S models. The upper limits on the signal strength are translated into constraints on $\sigma (\Pp\Pp \to \PH+\PX) {\mathcal{B}} (\PH \to \PaO \PaO)$ by scaling them with the theoretically predicted values of $\mathcal{B}^{2} (\PaO \to \PGt\PGt)$. The branching fraction, which depends on the model type, \maO, and $\tan\beta$, is calculated using the decay width expressions from Ref.~\cite{Haisch:2018kqx}. Figure~\ref{fig:modeldependentlimits2D} shows the observed and expected 95\% \CL upper limits on $\sigma (\Pp\Pp \to \PH+\PX) {\mathcal{B}} (\PH \to \PaO \PaO)$ obtained for the four types of 2HD+S models at benchmark values of $\tan\beta$, corresponding to scenarios where the $\PaO \to \PGt \PGt$ decay has a sizable branching fraction. Among the scenarios considered, the Type~III 2HD+S model with $\tan\beta = 2$ provides the most stringent limits across all \maO values between 4 and 15\GeV. The observed upper limits range from 0.010 at $\maO = 9\GeV$ to 0.35 at $\maO = 4\GeV$. For the Type~II model, for $\tan\beta = 5$, tight constraints between 0.013 and 0.092 are obtained for \maO up to 9\GeV. Above 9\GeV, the decay of \PaO to bottom quarks overwhelms the decay to tau leptons, making the analysis less sensitive. In the Type~I 2HD+S model, $\PaO$ couplings to fermions are independent of $\tan\beta$, so the upper limit depends only on \maO. The branching fraction to $\PGt\PGt$ is less enhanced since it has to compete with other decays, such as those to charm and bottom quark pairs. Thus, the analysis yields less stringent constraints on $\sigma (\Pp\Pp \to \PH+\PX) {\mathcal{B}} (\PH \to \PaO \PaO)$, ranging from 0.038 at 9\GeV to 29 at 4\GeV. For the Type~IV model, the analysis is only sensitive for $\tan\beta < 1$, since at higher values the decays to quarks dominate in the considered mass range. At $\tan\beta=0.5$, the analysis sets observed upper limits between 0.039 at 9\GeV and 29 at 4\GeV. 

The peak-like structures seen in the expected limits of Fig.~\ref{fig:modeldependentlimits2D} occur in \maO regions where quarkonium states, such as $\eta_\cPqc$ and $\eta_\PQb$, are found. In these regions, the mixing of \PaO-quarkonium states plays a crucial role, leading to a sudden increase in the hadronic decay width due to nonperturbative QCD effects. This results in a significant decrease of the branching fractions to unbound systems, such as $\PGt\PGt$. The mixing of \PaO with $\eta_c$ ($\eta_b$) is significantly amplified in scenarios where the coupling of \PaO to charm (bottom) quarks is enhanced. Further details of the \PaO-quarkonium mixing can be found in Refs.~\cite{Haisch:2018kqx, Baumgart:2012pj}. As a result of the mixing, the analysis fails to provide tight constraints in these mass regions. 

\begin{figure}[htbp!]
\centering
\includegraphics[width=0.49\textwidth]{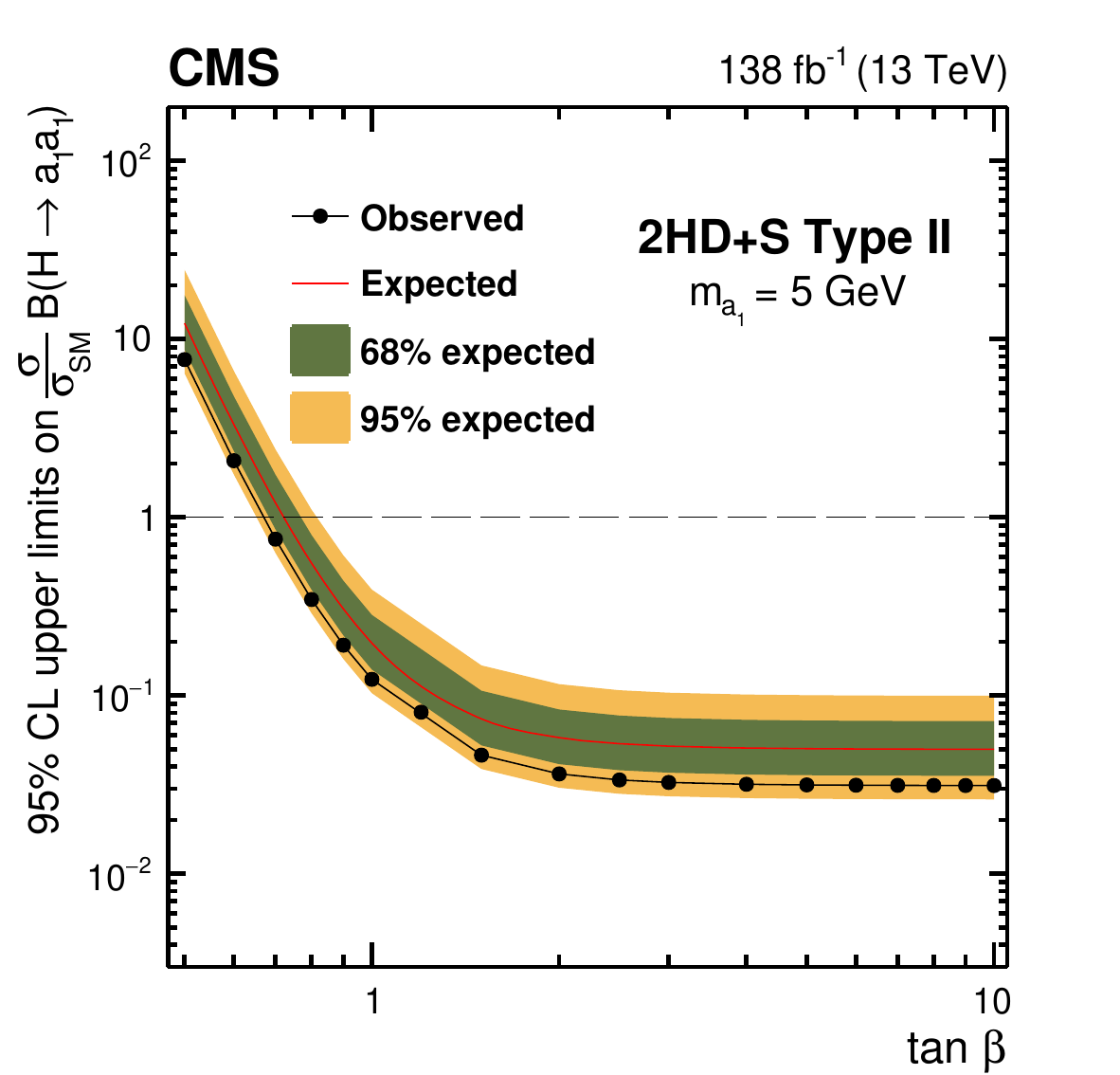}
\includegraphics[width=0.49\textwidth]{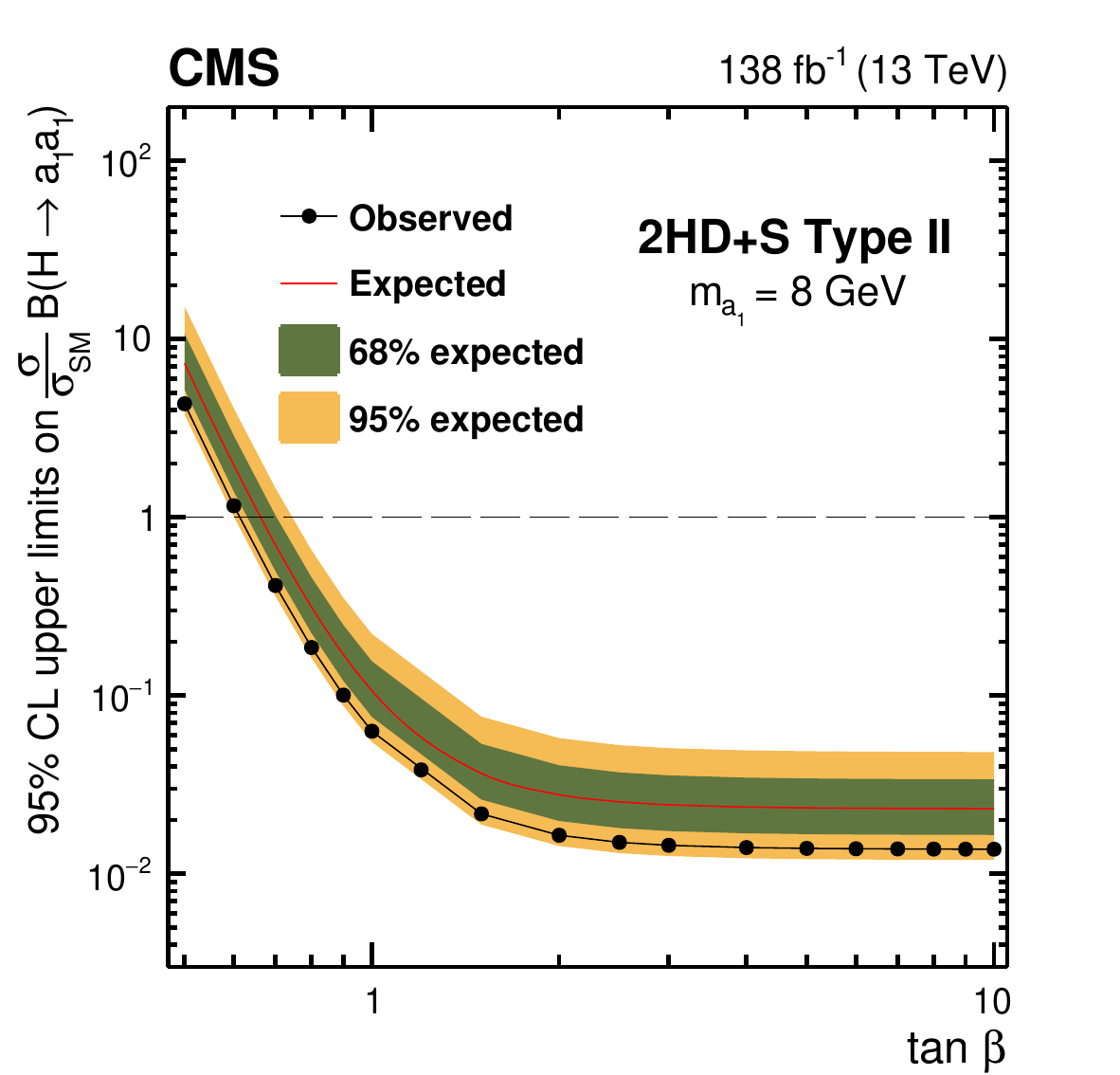}
\includegraphics[width=0.49\textwidth]{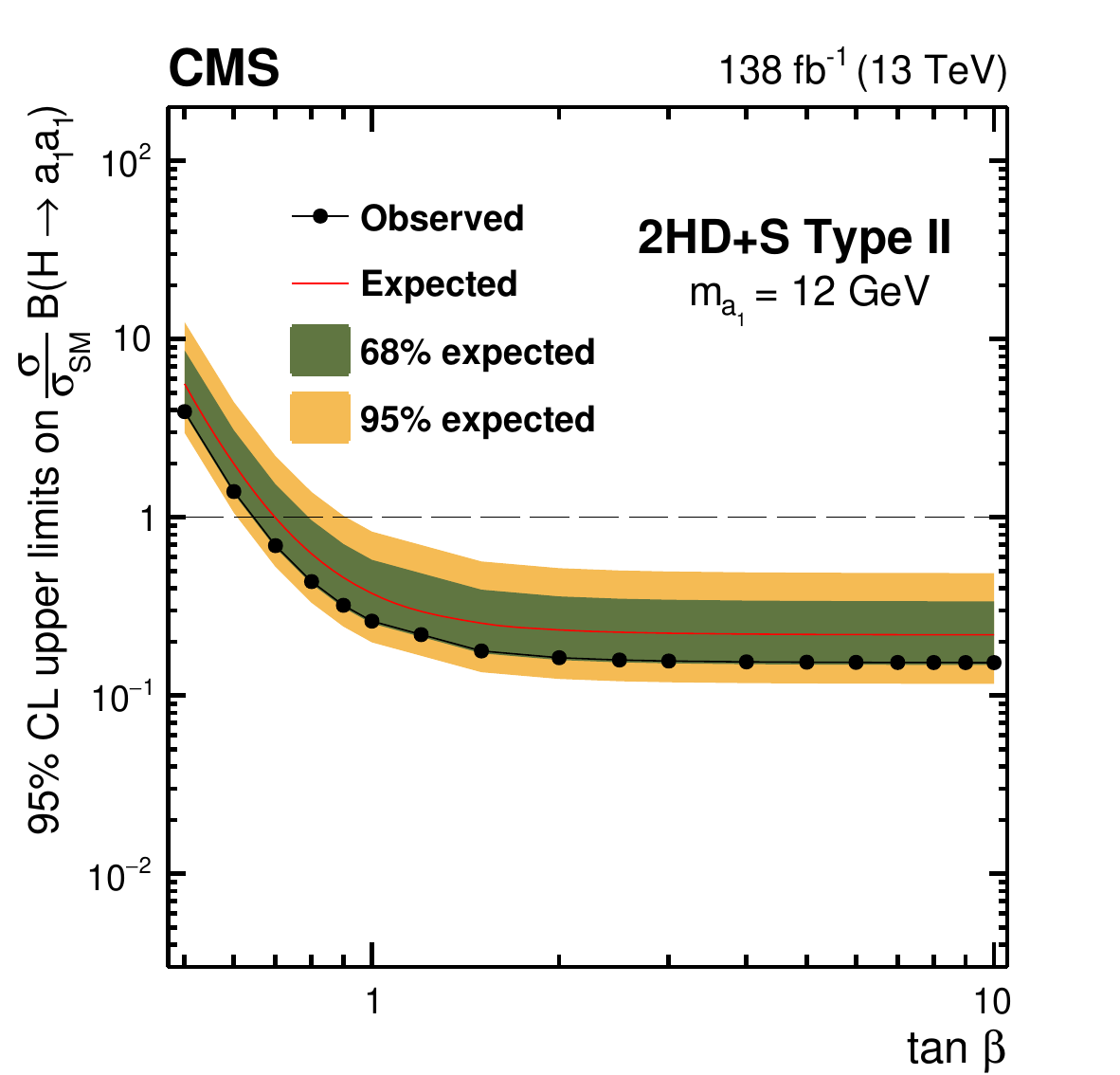}
\includegraphics[width=0.49\textwidth]{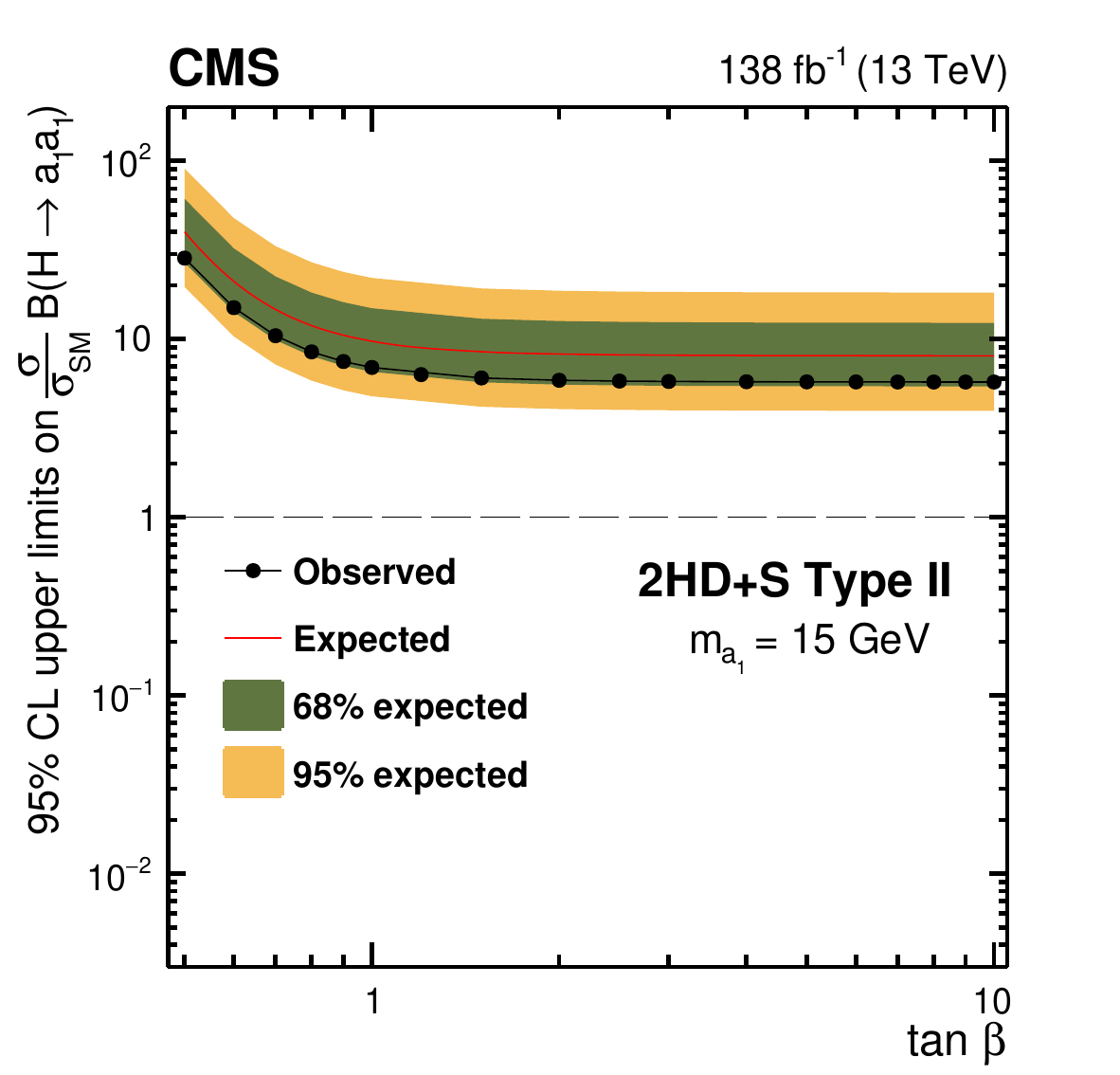}
\caption{
The observed (points) and expected (red line) 95\% \CL upper limits on $\sigma (\Pp\Pp \to \PH+\PX) {\mathcal{B}} (\PH \to \PaO \PaO)$, relative to $\sigma_\text{SM}$, as a function of $\tan\beta$ for the Type~II 2HD+S model with $\maO= 5\GeV$ (upper left), $\maO= 8\GeV$ (upper right), $\maO= 12\GeV$ (lower left), and $\maO= 15\GeV$ (lower right).
\label{fig:modeldependentlimits_tanbeta_2}
}
\end{figure}

\begin{figure}[hbpt!]
\centering
\includegraphics[width=0.49\textwidth]{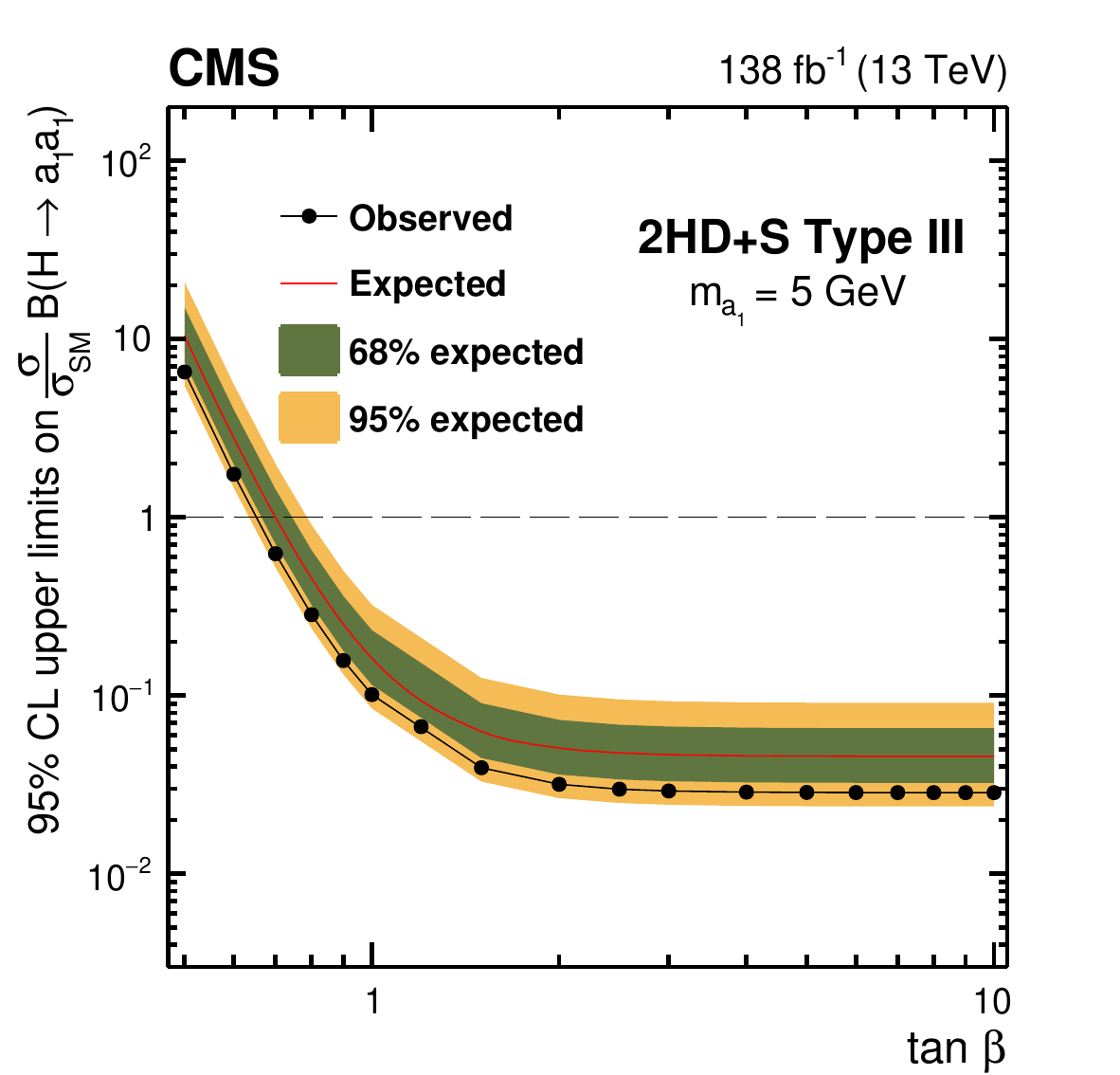}
\includegraphics[width=0.49\textwidth]{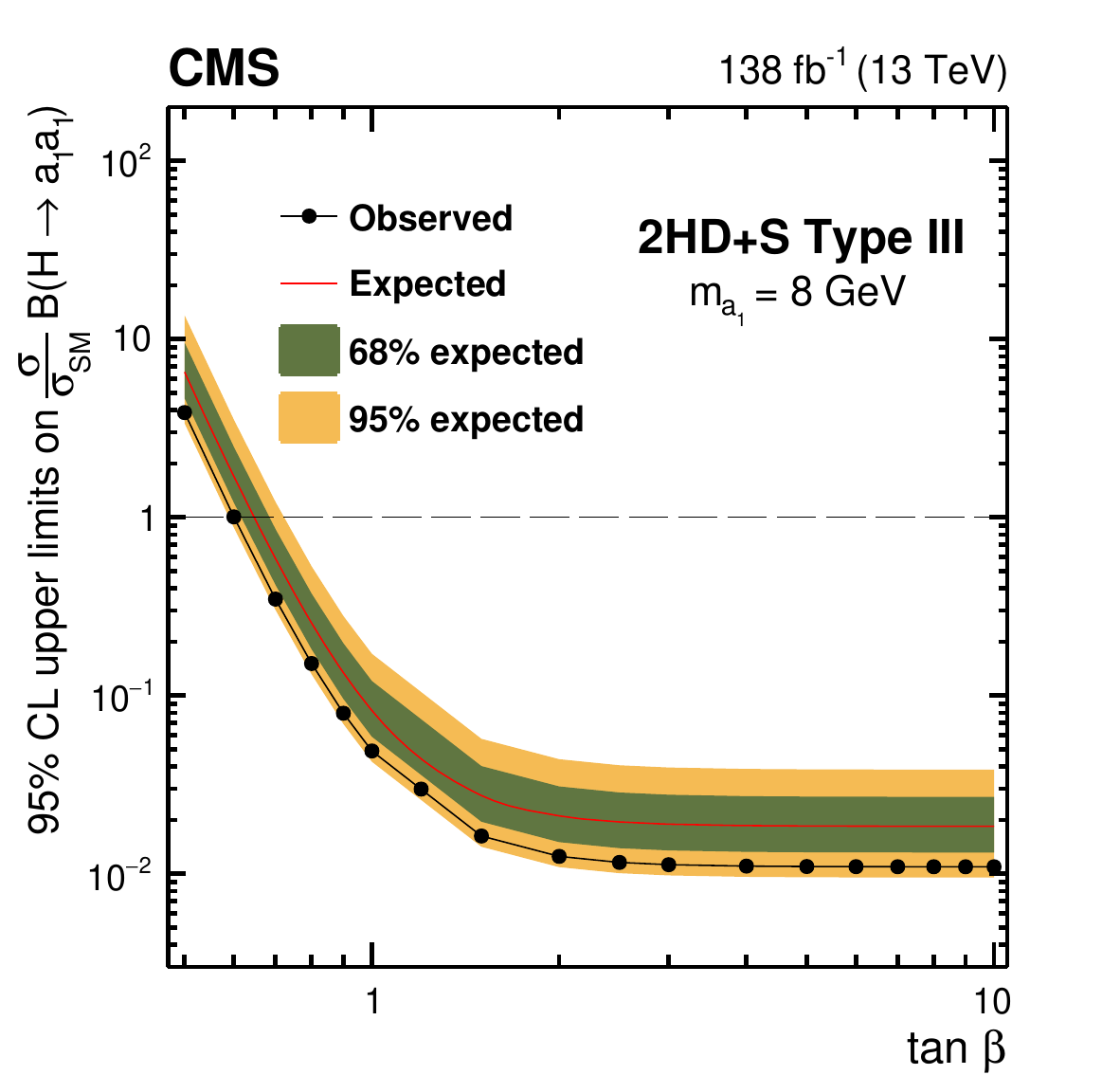}
\includegraphics[width=0.49\textwidth]{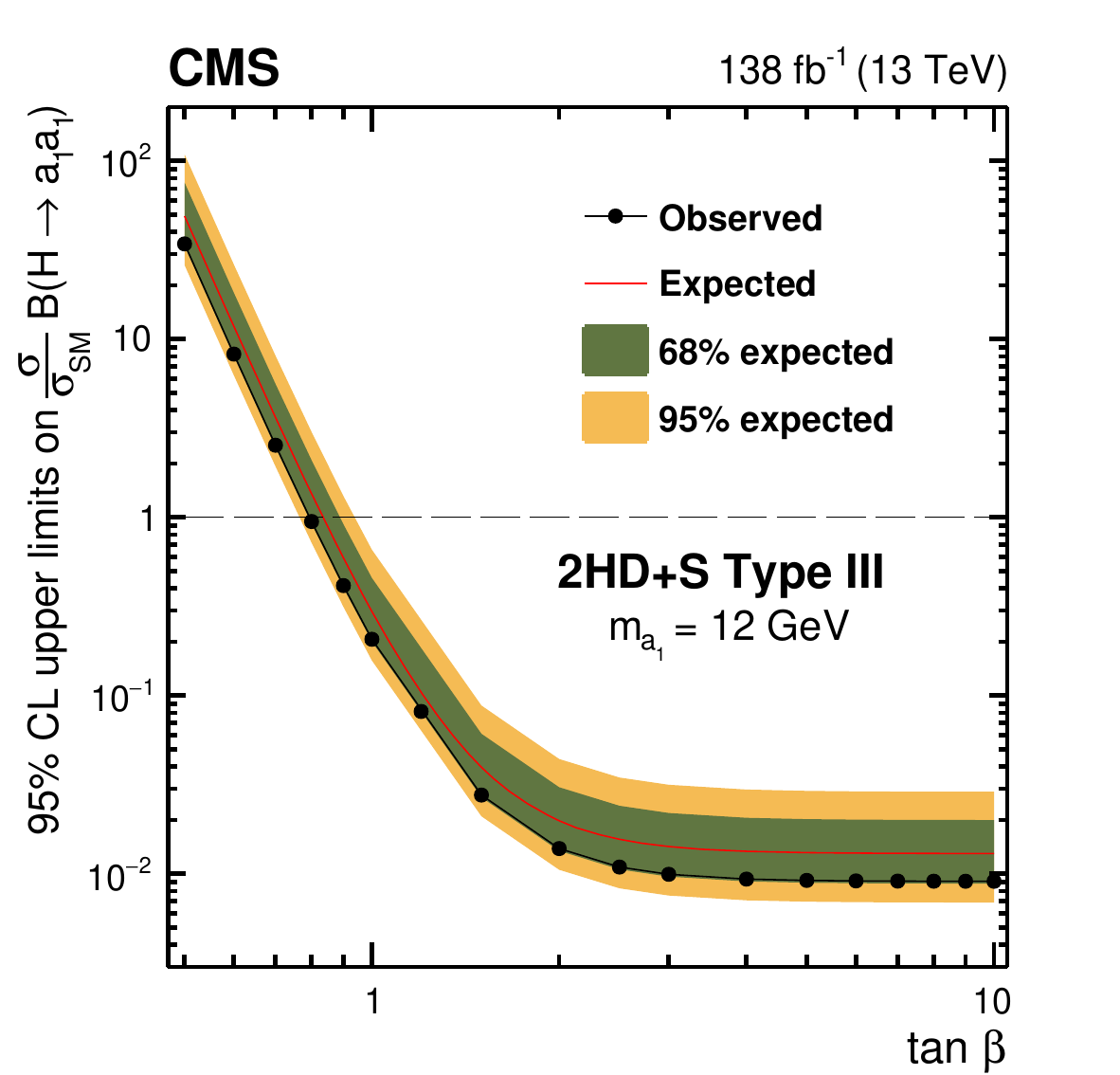}
\includegraphics[width=0.49\textwidth]{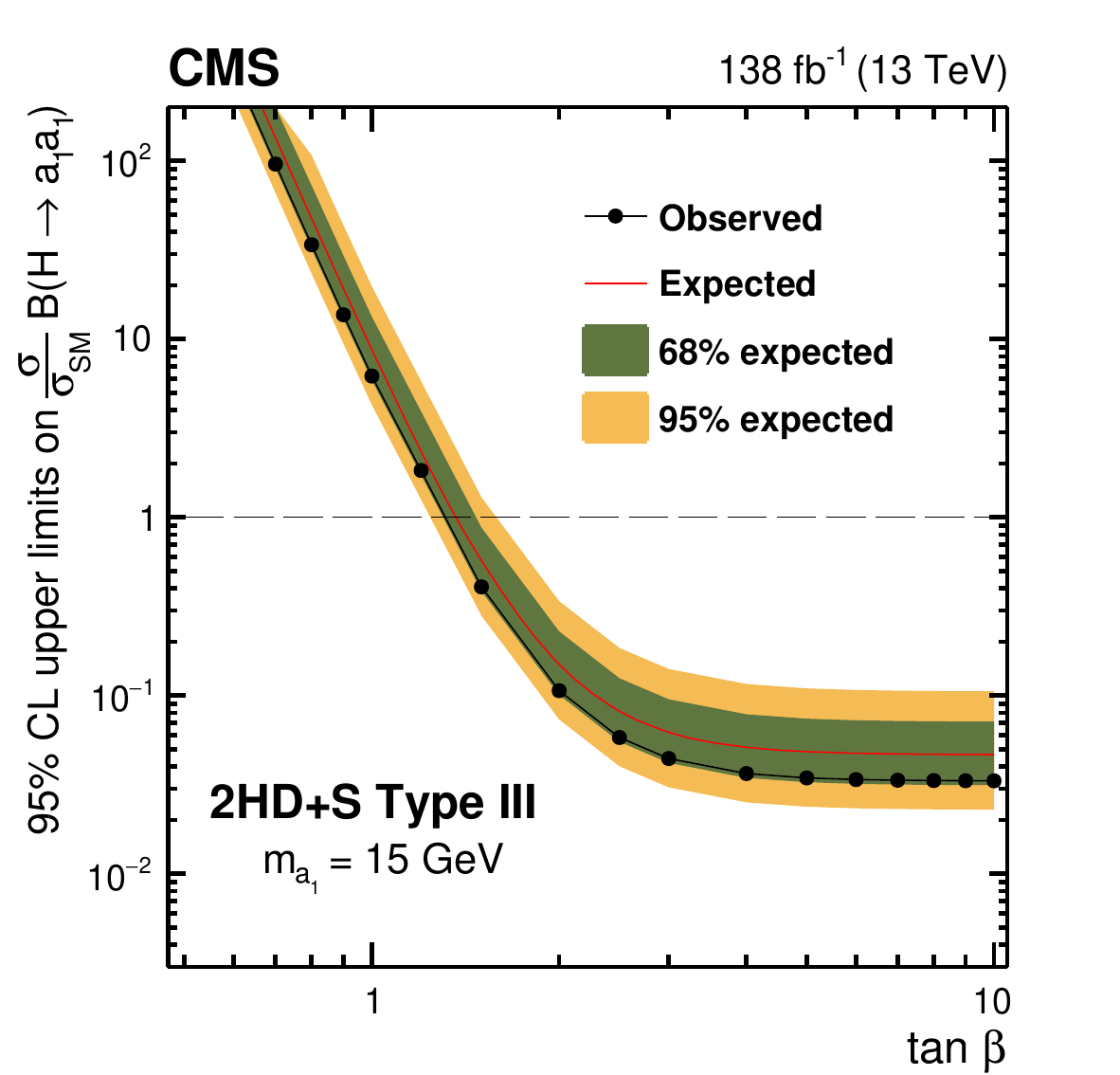}
\caption{
The observed (points) and expected (red line) 95\% \CL upper limits on $\sigma (\Pp\Pp \to \PH+\PX) {\mathcal{B}} (\PH \to \PaO \PaO)$, relative to $\sigma_\text{SM}$, as a function of $\tan\beta$ for the Type~III 2HD+S model with $\maO= 5\GeV$ (upper left), $\maO= 8\GeV$ (upper right), $\maO= 12\GeV$ (lower left), and $\maO= 15\GeV$ (lower right).
\label{fig:modeldependentlimits_tanbeta_3}
}
\end{figure}

Upper limits at 95\% \CL are also set on $\sigma (\Pp\Pp \to \PH+\PX) {\mathcal{B}} (\PH \to \PaO \PaO)$, relative to $\sigma_\text{SM}$, as a function of $\tan\beta$ for benchmark \PaO boson masses. Figures~\ref{fig:modeldependentlimits_tanbeta_2} and~\ref{fig:modeldependentlimits_tanbeta_3} present the limits obtained for Type~II and Type~III 2HD+S models, respectively, for $\maO= 5$, 8, 12, and 15\GeV. For both models, the analysis sets stringent constraints for $\tan\beta > 1$, where the coupling to tau leptons is enhanced. For $\tan\beta < 1$, decays of the \PaO boson to quarks dominate and suppress the $\PGt\PGt$ decay, resulting in weaker limits. The Type~III 2HD+S model provides the best constraints for $\tan\beta > 1$ across all considered \maO values. In the Type~II model, tight constraints are obtained for $5<\maO<8\GeV$, but the limits deteriorate for higher masses due to the enhanced Yukawa coupling to $\PQb$ quarks. The observed deterioration of the limits at lower and higher \PaO masses for both models can be directly related to the trends observed in the model-independent results. The enhanced $\PQb$ quark coupling plays an additional role in weakening the limits for higher masses in the Type~II model.

\section{Summary}
\label{Sec:summary}

A search for a pair of light pseudoscalar bosons (\PaO) produced in decays of the 125\GeV Higgs boson (\PH), $\PH\to\PaO\PaO$, in final states with two muons and two charged-particle tracks is presented. The search is performed using data from proton-proton collisions at a center-of-mass energy of 13\TeV, collected by the CMS experiment at the LHC between 2016 and 2018, and corresponding to an integrated luminosity of 138\fbinv. The analysis exploits the gluon-gluon fusion, vector boson fusion, and Higgs-strahlung production modes, and targets the $\PH\to \PaO \PaO \to 4\PGt$ and $2\PGm2\PGt$ decay channels. Masses of the \PaO boson ($\maO$) in the range 4--15\GeV are examined. No excess of data above the standard model (SM) background prediction is found. Upper limits on the product of the inclusive signal cross section and the branching fraction, $\sigma (\Pp\Pp \to \PH+\PX) {\mathcal{B}} (\PH \to \PaO \PaO) \mathcal{B}^2(\PaO \to \PGt\PGt)$, relative to the SM Higgs boson production cross section $\sigma_\text{SM}$, are set at 95\% confidence level (\CL) by combining the $4\PGt$ and $2\PGm2\PGt$ decay channels, assuming Yukawa-like couplings of \PaO to fermions. The observed limits range from 0.007 at $\maO=11\GeV$ to 0.079 at $\maO=4\GeV$. The expected limits in the absence of signal span from 0.011 at $\maO=11\GeV$ to 0.066 at $\maO=4\GeV$. The results are a significant improvement over the previous CMS analysis at 13\TeV~\cite{CMS:2019spf}, exceeding the anticipated improvement from the larger data sample alone. The sensitivity is enhanced by a factor of 2 to 4, depending on \maO, which can be attributed to the introduction of a veto for \cPqb-tagged jets and further optimization of the selection criteria targeting the $\PaO\to\PGt\PGt$ and $\PGm\PGm$ decays. The analysis also exceeds the sensitivity of a similar search performed by the ATLAS Collaboration in the same channel using a comparable amount of integrated luminosity~\cite{ATLAS:2025qyn}.

The results of the search are also interpreted in the context of various models with two Higgs doublets and an additional complex singlet field (2HD+S). For the Type~II 2HD+S scenario, realized in the next-to-minimal supersymmetric SM, 95\% \CL upper limits between 0.013 and 0.092 are set on $\sigma (\Pp\Pp \to \PH+\PX) {\mathcal{B}} (\PH \to \PaO \PaO)$, relative to $\sigma_\text{SM}$, for $4 < \maO < 9\GeV$ and $\tan\beta > 5$. The analysis sets the most stringent constraints to date for the Type~III 2HD+S scenario. Upper limits in the range 0.010--0.057 are obtained for probed mass hypotheses in the ranges $5 < \maO < 9\GeV$ and $12 < \maO < 14\GeV$ for $\tan\beta>2$.

\begin{acknowledgments}
We congratulate our colleagues in the CERN accelerator departments for the excellent performance of the LHC and thank the technical and administrative staffs at CERN and at other CMS institutes for their contributions to the success of the CMS effort. In addition, we gratefully acknowledge the computing centers and personnel of the Worldwide LHC Computing Grid and other centers for delivering so effectively the computing infrastructure essential to our analyses. Finally, we acknowledge the enduring support for the construction and operation of the LHC, the CMS detector, and the supporting computing infrastructure provided by the following funding agencies: SC (Armenia), BMBWF and FWF (Austria); FNRS and FWO (Belgium); CNPq, CAPES, FAPERJ, FAPERGS, and FAPESP (Brazil); MES and BNSF (Bulgaria); CERN; CAS, MoST, and NSFC (China); MINCIENCIAS (Colombia); MSES and CSF (Croatia); RIF (Cyprus); SENESCYT (Ecuador); ERC PRG, TARISTU24-TK10 and MoER TK202 (Estonia); Academy of Finland, MEC, and HIP (Finland); CEA and CNRS/IN2P3 (France); SRNSF (Georgia); BMBF, DFG, and HGF (Germany); GSRI (Greece); NKFIH (Hungary); DAE and DST (India); IPM (Iran); SFI (Ireland); INFN (Italy); MSIT and NRF (Republic of Korea); MES (Latvia); LMTLT (Lithuania); MOE and UM (Malaysia); BUAP, CINVESTAV, CONACYT, LNS, SEP, and UASLP-FAI (Mexico); MOS (Montenegro); MBIE (New Zealand); PAEC (Pakistan); MES and NSC (Poland); FCT (Portugal);  MESTD (Serbia); MICIU/AEI and PCTI (Spain); MOSTR (Sri Lanka); Swiss Funding Agencies (Switzerland); MST (Taipei); MHESI and NSTDA (Thailand); TUBITAK and TENMAK (T\"{u}rkiye); NASU (Ukraine); STFC (United Kingdom); DOE and NSF (USA).

\hyphenation{Rachada-pisek} Individuals have received support from the Marie-Curie program and the European Research Council and Horizon 2020 Grant, contract Nos.\ 675440, 724704, 752730, 758316, 765710, 824093, 101115353, 101002207, 101001205, and COST Action CA16108 (European Union); the Leventis Foundation; the Alfred P.\ Sloan Foundation; the Alexander von Humboldt Foundation; the Science Committee, project no. 22rl-037 (Armenia); the Fonds pour la Formation \`a la Recherche dans l'Industrie et dans l'Agriculture (FRIA-Belgium); the Beijing Municipal Science \& Technology Commission, No. Z191100007219010, the Fundamental Research Funds for the Central Universities, the Ministry of Science and Technology of China under Grant No. 2023YFA1605804, and the Natural Science Foundation of China under Grant No. 12061141002 (China); the Ministry of Education, Youth and Sports (MEYS) of the Czech Republic; the Shota Rustaveli National Science Foundation, grant FR-22-985 (Georgia); the Deutsche Forschungsgemeinschaft (DFG), among others, under Germany's Excellence Strategy -- EXC 2121 ``Quantum Universe" -- 390833306, and under project number 400140256 - GRK2497; the Hellenic Foundation for Research and Innovation (HFRI), Project Number 2288 (Greece); the Hungarian Academy of Sciences, the New National Excellence Program - \'UNKP, the NKFIH research grants K 131991, K 133046, K 138136, K 143460, K 143477, K 146913, K 146914, K 147048, 2020-2.2.1-ED-2021-00181, TKP2021-NKTA-64, and 2021-4.1.2-NEMZ\_KI-2024-00036 (Hungary); the Council of Science and Industrial Research, India; ICSC -- National Research Center for High Performance Computing, Big Data and Quantum Computing, FAIR -- Future Artificial Intelligence Research, and CUP I53D23001070006 (Mission 4 Component 1), funded by the NextGenerationEU program (Italy); the Latvian Council of Science; the Ministry of Education and Science, project no. 2022/WK/14, and the National Science Center, contracts Opus 2021/41/B/ST2/01369 and 2021/43/B/ST2/01552 (Poland); the Funda\c{c}\~ao para a Ci\^encia e a Tecnologia, grant CEECIND/01334/2018 (Portugal); the National Priorities Research Program by Qatar National Research Fund;  MICIU/AEI/10.13039/501100011033, ERDF/EU, "European Union NextGenerationEU/PRTR", and Programa Severo Ochoa del Principado de Asturias (Spain); the Chulalongkorn Academic into Its 2nd Century Project Advancement Project, and the National Science, Research and Innovation Fund via the Program Management Unit for Human Resources \& Institutional Development, Research and Innovation, grant B39G680009 (Thailand); the Kavli Foundation; the Nvidia Corporation; the SuperMicro Corporation; the Welch Foundation, contract C-1845; and the Weston Havens Foundation (USA).
\end{acknowledgments}

\bibliography{auto_generated} 
\cleardoublepage \appendix\section{The CMS Collaboration \label{app:collab}}\begin{sloppypar}\hyphenpenalty=5000\widowpenalty=500\clubpenalty=5000
\cmsinstitute{Yerevan Physics Institute, Yerevan, Armenia}
{\tolerance=6000
V.~Chekhovsky, A.~Hayrapetyan, V.~Makarenko\cmsorcid{0000-0002-8406-8605}, A.~Tumasyan\cmsAuthorMark{1}\cmsorcid{0009-0000-0684-6742}
\par}
\cmsinstitute{Institut f\"{u}r Hochenergiephysik, Vienna, Austria}
{\tolerance=6000
W.~Adam\cmsorcid{0000-0001-9099-4341}, J.W.~Andrejkovic, L.~Benato\cmsorcid{0000-0001-5135-7489}, T.~Bergauer\cmsorcid{0000-0002-5786-0293}, S.~Chatterjee\cmsorcid{0000-0003-2660-0349}, K.~Damanakis\cmsorcid{0000-0001-5389-2872}, M.~Dragicevic\cmsorcid{0000-0003-1967-6783}, P.S.~Hussain\cmsorcid{0000-0002-4825-5278}, M.~Jeitler\cmsAuthorMark{2}\cmsorcid{0000-0002-5141-9560}, N.~Krammer\cmsorcid{0000-0002-0548-0985}, A.~Li\cmsorcid{0000-0002-4547-116X}, D.~Liko\cmsorcid{0000-0002-3380-473X}, I.~Mikulec\cmsorcid{0000-0003-0385-2746}, J.~Schieck\cmsAuthorMark{2}\cmsorcid{0000-0002-1058-8093}, R.~Sch\"{o}fbeck\cmsAuthorMark{2}\cmsorcid{0000-0002-2332-8784}, D.~Schwarz\cmsorcid{0000-0002-3821-7331}, M.~Sonawane\cmsorcid{0000-0003-0510-7010}, W.~Waltenberger\cmsorcid{0000-0002-6215-7228}, C.-E.~Wulz\cmsAuthorMark{2}\cmsorcid{0000-0001-9226-5812}
% ...
\par}
\vskip\cmsinstskip
\dag:~Deceased\\
$^{1}$Also at Yerevan State University, Yerevan, Armenia\\
$^{2}$Also at TU Wien, Vienna, Austria\\
% ...
\end{sloppypar}
\end{document}